\documentclass[10pt,preprint]{aastex}
\usepackage{psboxit}

\PScommands
\newcommand{\graybox}[1]{\psboxit{box 0.7 setgray fill}{\spbox{#1}}}

\newcommand{\latin}[1]{{#1}}

\newcommand{\eg}{\latin{e.g.}}

\newcommand{\Sersic}{S\'ersic}

\def\simless{\mathbin{\lower 3pt\hbox
	{$\,\rlap{\raise 5pt\hbox{$\char'074$}}\mathchar"7218\,$}}} 
\def\simgreat{\mathbin{\lower 3pt\hbox
	{$\,\rlap{\raise 5pt\hbox{$\char'076$}}\mathchar"7218\,$}}} 

\newcommand{\band}[2]{\ensuremath{^{#1}\!{#2}}}
\newcommand{\Vmax}{\ensuremath{V_\mmax}}
\newcommand{\mmax}{\ensuremath{\mathrm{max}}}

\newcounter{thefigs}
\newcommand{\fignum}{\arabic{thefigs}}

\newcounter{thetabs}

\newcounter{address}

\slugcomment{Submitted to \apj}

\shortauthors{Blanton {\it et al.} (2002)}
\shorttitle{Extremely low luminosity galaxies}

\begin{document}
 

\title{The properties and luminosity function \\
of extremely low 
luminosity galaxies\altaffilmark{1}}



\author{
Michael R. Blanton\altaffilmark{\ref{NYU}}, 
Robert H. Lupton\altaffilmark{\ref{Princeton}},
David J. Schlegel\altaffilmark{\ref{Princeton}},
Michael A. Strauss\altaffilmark{\ref{Princeton}},
J. Brinkmann\altaffilmark{\ref{APO}},
Masataka Fukugita\altaffilmark{\ref{CosmicRay}}, and
Jon Loveday\altaffilmark{\ref{Sussex}}
}

\altaffiltext{1}{Based on observations obtained with the
Sloan Digital Sky Survey\label{SDSS}} 
\setcounter{address}{2}
\altaffiltext{\theaddress}{
\stepcounter{address}
Center for Cosmology and Particle Physics, Department of Physics, New
York University, 4 Washington Place, New
York, NY 10003
\label{NYU}}
\altaffiltext{\theaddress}{
\stepcounter{address}
Princeton University Observatory, Princeton,
NJ 08544
\label{Princeton}}
\altaffiltext{\theaddress}{
\stepcounter{address}
Apache Point Observatory,
2001 Apache Point Road,
P.O. Box 59, Sunspot, NM 88349-0059
\label{APO}}
\altaffiltext{\theaddress}{
\stepcounter{address}
Institute for Cosmic Ray Research, University of
Tokyo, Midori, Tanashi, Tokyo 188-8502, Japan
\label{CosmicRay}}
\altaffiltext{\theaddress}{
\stepcounter{address}
Sussex Astronomy Centre,
University of Sussex,
Falmer, Brighton BN1 9QJ, UK
\label{Sussex}}

\begin{abstract}
We examine a sample of low redshift ($10<d<150$ $h^{-1}$ Mpc) field
galaxies including galaxies with luminosities as low as $M_r - 5
\log_{10} h \sim -12.5$, selected from the Sloan Digital Sky Survey
Data Release 2 (SDSS). The sample is unique in containing galaxies of
extremely low luminosities in a wide range of environments, selected
with uniform and well-understood criteria.  We present the luminosity
function as well as the broad-band properties of low luminosity
galaxies in this sample. A Schechter function is an insufficient
parameterization of the $r$-band luminosity function; there is an
upturn in the slope for $M_r - 5 \log_{10} h > -18$. The resulting
slope at low luminosities in this sample is $\alpha_2 \sim
-1.3$. However, we almost certainly miss a large number of galaxies at
very low luminosities due to low surface brightness selection effects,
and we estimate that the true low luminosity slope may be as steep as
or steeper than $\alpha_2 \sim -1.5$. The results here are consistent
with previous SDSS results and, in the $g$-band, roughly consistent
with the results of the Two degree Field Galaxy Redshift
Survey. Extremely low luminosity galaxies are predominantly low
surface brightness, exponential disks, the majority of which are red.
\end{abstract}


\section{Motivation}
\label{motivation}

Extremely low luminosity galaxies in the field are not a well-studied
population, because they are difficult to find except in deep,
wide-field surveys. On the other hand, one of the major problems
facing models of galaxy formation is why cold dark matter models
predict larger numbers of low mass galaxies than observed, either in
the field or in the halos of larger galaxies. Thus, understanding the
properties of the lowest luminosity galaxies (which are presumably
also very low mass) might shed light on some of the mysteries
surrounding the production of galaxies in low mass halos.

Most of the observational work regarding low luminosity galaxies has
concentrated on galaxies in clusters. The Virgo cluster photometric
sample of low luminosity galaxies (\citealt{binggeli85b}), for which
the background contamination is low, indicates that the low luminosity
galaxies in clusters tend to be red and exponential in profile shape
(e.g. \citealt{barazza03a}).  The low luminosity slope of the
luminosity function in the Virgo cluster appears to be $\alpha \sim
-1.3$. A number of other purely photometric studies have found
similarly steep slopes, usually as steep as or steeper than $\alpha
\sim -1.4$ (\citealt{depropris95a,trentham98a,secker97a}), and
possibly a variation of the low luminosity slope depending on the
cluster (\citealt{driver98a}). However, it is worth noting that
deriving luminosity functions from photometry alone requires
background subtraction, an uncertain procedure which may bias one's
estimates towards steep slopes (\citealt{valotto01a}).

Recently, \citet{deady02a} have studied the properties of
spectroscopically confirmed dwarf galaxies in the Fornax cluster. They
report that most of the low luminosity galaxies in that cluster are dE
galaxies that lie on the optical color-magnitude and magnitude-surface
brightness relationships for dwarf galaxies. Notably, they find a
population of ultra compact dwarfs (UCDs) which tend to be very red
and (as their name suggests) small even for their luminosity.

The other location in which astronomers have studied very low
luminosity galaxies is of course in the Local Group
(\citealt{mateo98a}). It is in this environment that we have the best
constraints on the luminosity function at very low luminosities
(\citealt{vandenbergh92a, trentham05a}). In the Local Group, ``late
type'' dwarfs (spirals or irregulars) tend to be blue, though a large
fraction ($\sim 50\%$) of the known dwarfs ($M_V > -14$) are ``early
type'' (dwarf spheroidals and ellipticals) and have red colors typical
of high luminosity galaxies.  In the past few years, there have been
several new detections of low luminosity objects in the Local Group
(\citealt{ibata95a, armandroff98a, armandroff99a, whiting99a,
zucker04a, willman04a}).

There are a number of efforts to collect data on low luminosity
galaxies in other areas. \citet{impey96a} have searched for low
luminosity galaxies nearby by selecting low surface brightness
galaxies.  More recently, \citet{karachentsev04a} released a catalog
of nearby galaxies selected from the literature, from eyeball
inspection of photographic plates, from HI surveys, and from infrared
surveys. Again, many of the galaxies were found by selecting low
surface brightness objects which were followed up in the optical or
radio. The resulting catalog contains galaxies as low luminosity as
$M_B \sim -8$. \citet{gildepaz03a} have defined a set of ``blue
compact dwarf'' (BCD) galaxies, having culled these from a number of
different catalogs with varying selection criteria. A new very low
luminosity galaxy ($M_I \sim -10$) has been recently discovered in the
field (\citealt{pasquali04a}). There are also viable candidates (as
yet unconfirmed) in small groups (\citealt{flint01a}). The HI Parkes
All-Sky Survey (HIPASS;
\citealt{meyer04a}) have a neutral-gas selected sample which includes
many objects with low optical luminosities and surface brightnesses
(West et al. in preparation).

Previous large field galaxy redshift surveys have provided samples of
low luminosity galaxies selected without regard to environment: for
example, the Center for Astrophysics redshift survey
(\citealt{huchra83a}), the Las Campanas Redshift Survey
(\citealt{shectman96a}), the IRAS Point Source Catalog redshift survey
(\citealt{saunders00a}), and the Two degree Field Galaxy Redshift
Survey (2dFGRS; \citealt{colless01a}). The Sloan Digital Sky Survey
(SDSS; \citealt{abazajian04a}) represents an improvement on these and
other previous surveys. It has deep surface brightness limits in its
photometry, covers a wide area of sky, and provides highly complete
follow-up spectroscopy. On the other hand, its photometric and
spectroscopic pipelines are not optimized for finding such low
redshift objects, resulting in problems in creating a reliable low
redshift catalog. In \citet{blanton04a} we created such a catalog
using extra processing and careful inspection of galaxies in the SDSS
catalog. The result is not perfect; in particular, there is clearly
incompleteness at the luminous, nearby end and at low surface
brightness. However, the selection process, which is homogeneous,
fairly well-defined, and unbiased with respect to environment, yields
a catalog whose properties we can compare to theoretical predictions.

Our focus here is not on studying a necessarily {\it complete} census
of nearby galaxies --- impossible in any case given the current sky
coverage of the SDSS --- but instead on describing the properties of a
well-defined sample. For that reason, we have not included every
object with a known (low) redshift in the catalog, but only those
whose selection criteria we understand --- that is, those in the SDSS
spectroscopic survey.

Our sample is publicly distributed as part of the NYU Value-Added
Galaxy Catalog (\citealt{blanton04a}). As described below, that
catalog includes all of the parameters used to make the figures in
this paper, including atlas images.

In this paper we present the catalog and a basic description of their
properties.  Section \ref{data} describes the SDSS, how we have
selected our galaxies, and what properties we have measured. Section
\ref{surface} investigates the surface brightness completeness of our
sample. Section \ref{properties} describes qualitatively the
distribution in space of these galaxies, the distribution of their
basic photometric properties relative to those of more luminous
galaxies, and their luminosity function.  Section
\ref{summary} summarizes our results.

Where necessary, we have assumed cosmological parameters $\Omega_0 =
0.3$, $\Omega_\Lambda = 0.7$, and $H_0 = 100$ $h$ km s$^{-1}$
Mpc$^{-1}$ (with $h=1$). All magnitudes in this paper are
$K$-corrected to rest-frame bandpasses using the method of
\citet{blanton03b}, unless otherwise specified.  Because of the small
range of look-back times in our sample (a maximum of around 700 Myr),
we do not evolution-correct any of our magnitudes.

\section{Data}
\label{data}

\subsection{SDSS}

The SDSS is taking $ugriz$ CCD imaging of $10^4~\mathrm{deg^2}$ of the
Northern Galactic sky, and, from that imaging, selecting $10^6$
targets for spectroscopy, most of them galaxies with
$r<17.77~\mathrm{mag}$ \citep[\eg,][]{gunn98a,york00a,abazajian03a}.
Automated software performs all of the data processing: astrometry
\citep{pier03a}; source identification, deblending and photometry
\citep{lupton01a}; photometricity determination \citep{hogg01a};
calibration \citep{fukugita96a,smith02a}; spectroscopic target
selection \citep{eisenstein01a,strauss02a,richards02a}; spectroscopic
fiber placement \citep{blanton03a}; and spectroscopic data reduction.
Descriptions of these pipelines also exist in \citet{stoughton02a}.
An automated pipeline called {\tt idlspec2d} measures the redshifts
and classifies the reduced spectra (Schlegel et al., in preparation).

The spectroscopy has small incompletenesses coming primarily from (1)
galaxies missed because of mechanical spectrograph constraints
\citep[6~percent;][]{blanton03a}, which leads to a slight
under-representation of high-density regions, and (2) spectra in which
the redshift is either incorrect or impossible to determine
($<1$~percent).  In this context, we note that the mechanical
constraints are due to the fact that fibers cannot be placed more
closely than 55$''$; when two or more galaxies have a separation
smaller than this distance, one member is chosen independent of its
magnitude or surface brightness. Thus, this incompleteness does not
bias the sample with respect to luminosity. In addition, there are
some galaxies ($\sim 1$~percent) blotted out by bright Galactic stars,
but this incompleteness should be uncorrelated with galaxy properties.

\subsection{NYU-VAGC}

For the purposes of computing large-scale structure and galaxy
property statistics, we have assembled a subsample of SDSS galaxies
known as the NYU Value Added Galaxy Catalog (NYU-VAGC;
\citealt{blanton04a}). One of the products of that catalog is a low
redshift catalog. Here we use the version of that catalog
corresponding to the SDSS Data Release 2 (DR2;
\citealt{abazajian04a}). The reader can obtain this catalog at our web
site\footnote{{\tt http://sdss.physics.nyu.edu/vagc}}; the low
redshift catalog is referred to on those pages as ``{\tt lowz}.''

The low redshift catalog has a number of important features which are
useful in the study of low luminosity galaxies. Most importantly:
\begin{enumerate}
\item We have checked by eye all of the images and spectra of low
luminosity ($M_r > -15$) or low redshift ($z<0.01$) galaxies in the
NYU-VAGC. Most significantly, we have trimmed those which are ``flecks''
incorrectly deblended out of bright galaxies; for some of these cases,
we have been able to replace the photometric measurements with the
measurements of the parents. For a full description of our checks, see
\citet{blanton04a}, in which we conclude that we are correctly
treating the photometry for $>90\%$ of the galaxies for which we have
spectroscopy.
\item The deblending algorithm has improved over time; we targeted the
spectroscopy using the original, inferior reductions, but have since
rereduced the images with a new version of the software. For galaxies
which were shredded in the target version of the deblending, the
spectra are often many arcseconds away from the nominal centers of the
galaxy in the latest version of the photometric reductions. We have
used the new version of the deblending to decide whether these
(otherwise non-matched spectra) should be associated with the galaxy
in the best version.
\item We have estimated the distance to low redshift objects by
correcting for peculiar velocities using the \citet{willick97a} model
of the local velocity field based on the IRAS 1.2 Jy redshift survey
(\citealt{yahil91a}) density field (using $\beta = 0.5$), and
propagated the uncertainties in distance into uncertainties in
absolute magnitude.
\end{enumerate}

Other general features of the NYU-VAGC low redshift catalog which we
make use of here are:
\begin{enumerate}
\item  As part of NYU-VAGC, we fit the radial galaxy profile using a simple
\Sersic\ measurement (\citealt{sersic68a}), accounting for seeing. The
form of the \Sersic\ profile is:
\begin{equation}
\label{sersic}
I(r) = A \exp\left[ - (r/r_0)^{1/n} \right].
\end{equation}
The fitting procedure is fully described in \citet{blanton03q}.
\item We estimate the maximum volume in which we could have observed
each galaxy, accounting for the flux limits and completeness of the
survey.
\item We calculate $K$-corrected and Galactic extinction corrected
(\citealt{schlegel98a}) absolute magnitudes in the $ugrizJHK_s$
passbands. The $JHK_s$ passband measurements are taken from the
Two-Micron All Sky Survey Extended Source Catalog (2MASS XSC).  All
magnitudes in this paper are $K$-corrected to rest-frame bandpasses
using the method of \citet{blanton03b}.  We do not evolution-correct
any of the magnitudes because of the small range of redshifts in our
sample.
\end{enumerate}

All surface brightnesses referred to in this paper are $r$-band
Petrosian half-light surface brightnesses $\mu_{50,r}$, unless
otherwise specified. 

We select galaxies for the catalog in the range $10 < d < 150$
$h^{-1}$ Mpc. The total completeness-weighted effective area of the
sample is 2220.9 square degrees. Thus, the volume of the sample is
$7.6\times 10^5$ $h^{-3}$ Mpc$^3$.  The catalog contains 28,089
galaxies.

\begin{figure}
\figurenum{\fignum a}
\plotone{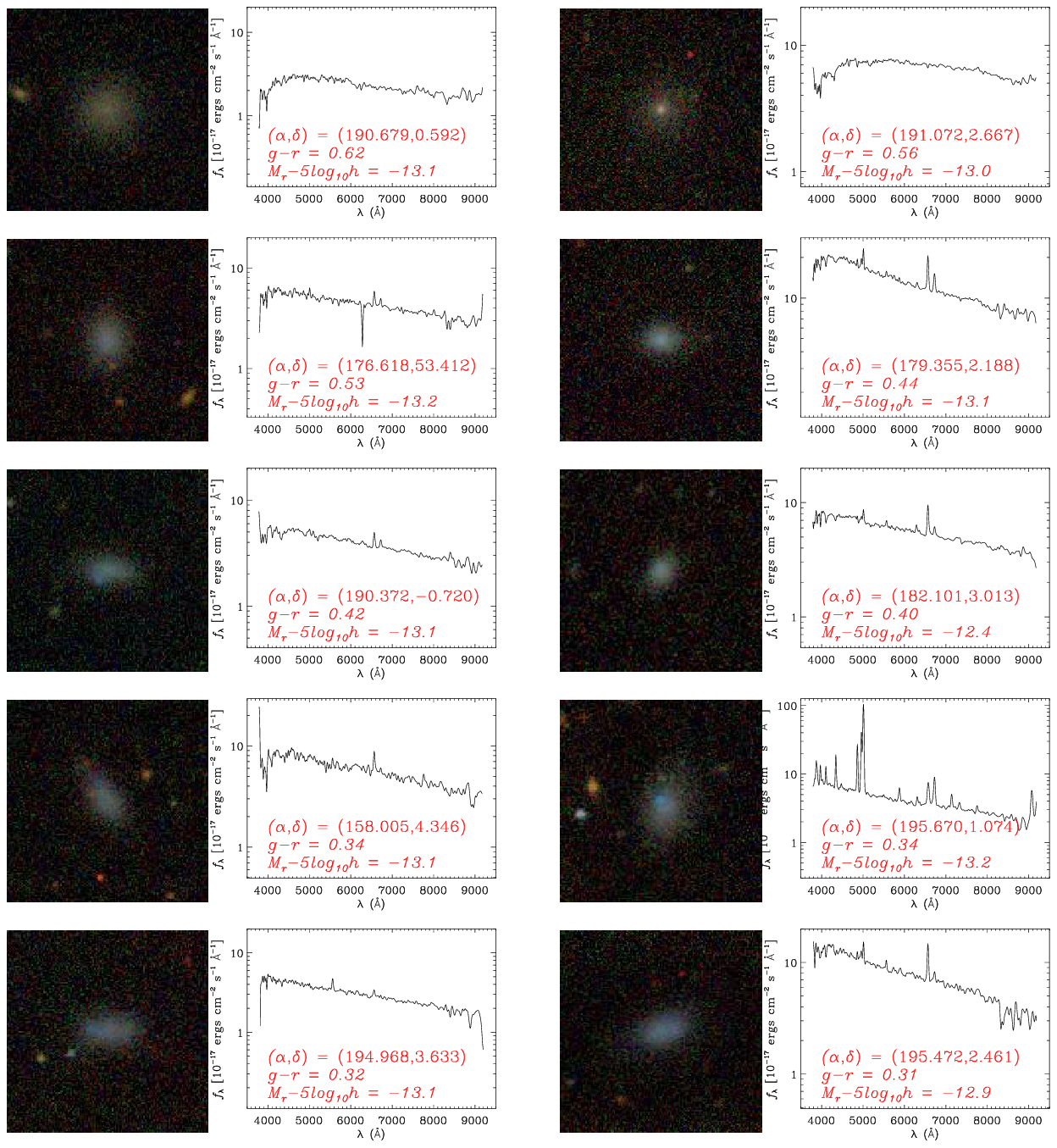}
\caption{\label{lli_1} SDSS $gri$ images and 3$''$ fiber spectra for
  the 20 lowest luminosity galaxies in our sample (all of which $M_r +
  5\log_{10} h > -13.2$). Each image is 57$''$ on a side. The spectrum
  in each case is associated with the center of the image. We have
  sorted them by $g-r$ color, starting with the reddest. }
\end{figure}

\begin{figure}
\figurenum{\fignum b}
\plotone{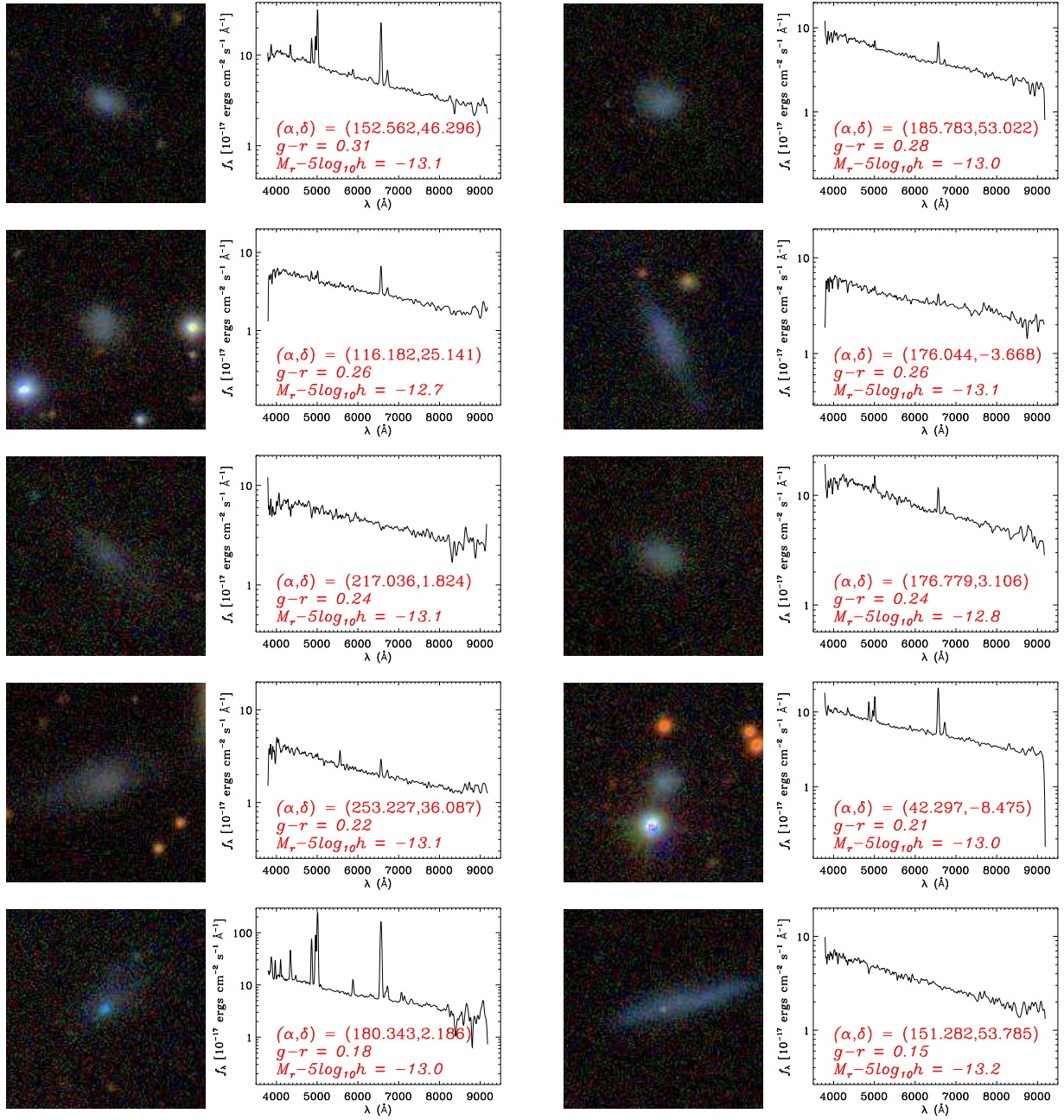}
\caption{\label{lli_2} Continuation of Figure \ref{lli_1}.}
\end{figure}

Figures \ref{lli_1} and \ref{lli_2} show the SDSS $gri$ images and
$3''$ fiber spectra for the 20 lowest luminosity galaxies in our
sample. The color images are prepared using the method of
\citet{lupton04a}. They are sorted by color, reddest to bluest. These
low luminosity galaxies tend to be blue, unconcentrated, have very
little structure, and no noticeable dust lanes. These properties are
consistent with previous studies of faint galaxies ({\it e.g.},
\citealt{marzke97a, bromley98a, madgwick02a}). They are mostly
emission-line galaxies. Among these 20 galaxies, only the reddest two
would be classified as dEs, according to their colors and spectra (one
of these two is nucleated).

As noted in the introduction, the advantage of our catalog is its
well-defined selection criteria. For comparison, the more local
catalog of \citet{karachentsev04a} (limited roughly at 7 $h^{-1}$ Mpc)
includes a large number of objects less luminous than those we study
here, most of them closer than 2.5 $h^{-1}$ Mpc. In addition, their
catalog covers the entire sky, allowing them to examine very local
structures, while ours is restricted to the SDSS area. Finally,
because they include HI-discovered galaxies and due to their efforts
at spectroscopic follow-up on low surface brightness objects, they
include some galaxies of much lower surface brightness than we include
here.  However, a disadvantage of their catalog is that it appears to
have been selected using very heterogeneous criteria, making certain
statistical studies difficult.

\section{Surface brightness completeness}
\label{surface}

Low luminosity galaxies tend to be low surface brightness. For that
reason, it is important to understand the surface brightness
incompleteness of our catalog. There are three aspects of this
incompleteness. First, in Section \ref{ph_fgot}, we determine the
completeness of the most recent version of the SDSS photometric
catalog by testing the software on simulated data. In Section
\ref{dusttest}, we test this completeness empirically. Second, in Section
\ref{ti_fgot}, we determine the completeness of the SDSS spectroscopic
targeting with respect to the photometric catalog. Third, in Section
\ref{sp_fgot}, we determine the redshift completeness of the SDSS
spectroscopy for targets which have been observed. Section
\ref{fgot}, summarizes the effects of surface brightness on
selection.

\subsection{The photometric catalog}
\label{ph_fgot}

We have examined the surface brightness completeness of the SDSS
photometric survey using a fake data pipeline (also described in
\citealt{blanton03q}).

For the current project we create 2500 fake galaxies (each with one
image in each of the five SDSS bands) with exponential profiles, blue
colors typical of low surface brightness galaxies, of varying axis
ratios, and of varying half-light surface brightnesses and magnitudes,
in the ranges $18 < \mu_{50,r} < 26$ and $14 < m_r < 17.5$. The faint
limit here is close to but slightly brighter than the SDSS
spectroscopic limit of 17.77; as this limit is almost 100 times
brighter than the photometric detection limit, the 30\% difference
between 17.5 and 17.77 is not important to our results.  We distribute
the fake stamps onto a random set of locations on the sky covered by
SDSS fields. For each observation of an object in an SDSS field in
each band, we convert the fake stamps to SDSS raw data units, convolve
with the estimated seeing from the SDSS photometric pipeline, and add
Poisson noise using the estimates of the gain. We add the resulting
image to the real SDSS raw data, including the tiny effects of
nonlinearity in the response and the less tiny flat-field variation as
a function of column on the chip. We run the SDSS photometric pipeline
{\tt photo} (version {\tt v5\_4\_25}, essentially that used in the
SDSS DR2 release) on the resulting set of images to extract and
measure objects.  This procedure thus includes the effects of seeing,
noise, and sky subtraction.

Using the resulting photometric catalog, we create a fake version of
the value-added catalog of \citet{blanton04a}, resolving multiple
observations and imposing the selection criteria in exactly the same
way as for the full catalog, including star-galaxy separation and all
of the other SDSS Main sample criteria used for spectroscopic target
selection (\citealt{strauss02a}). These criteria include an {\it
explicit} half-light surface brightness limit at $\mu_{50,r} =
24.5$. These steps ensure that the fake photometric sample is selected
in the same manner as the real one in the NYU-VAGC.

Finally, we compare the resulting catalog to the input catalog in
order to evaluate our completeness to surface brightness selection
effects. We require that the output object have a center within 5$''$
and a Petrosian magnitude within 0.5 mag of the input object (since
usually such a large magnitude discrepancy indicates that a separate,
nearby object has been targeted).  We should note here that in the
actual catalog we have reinstated some objects by hand. We have not
done so in the fake catalog, meaning that we slightly underestimate
our completeness.

\begin{figure}
\figurenum{\fignum}
\plotone{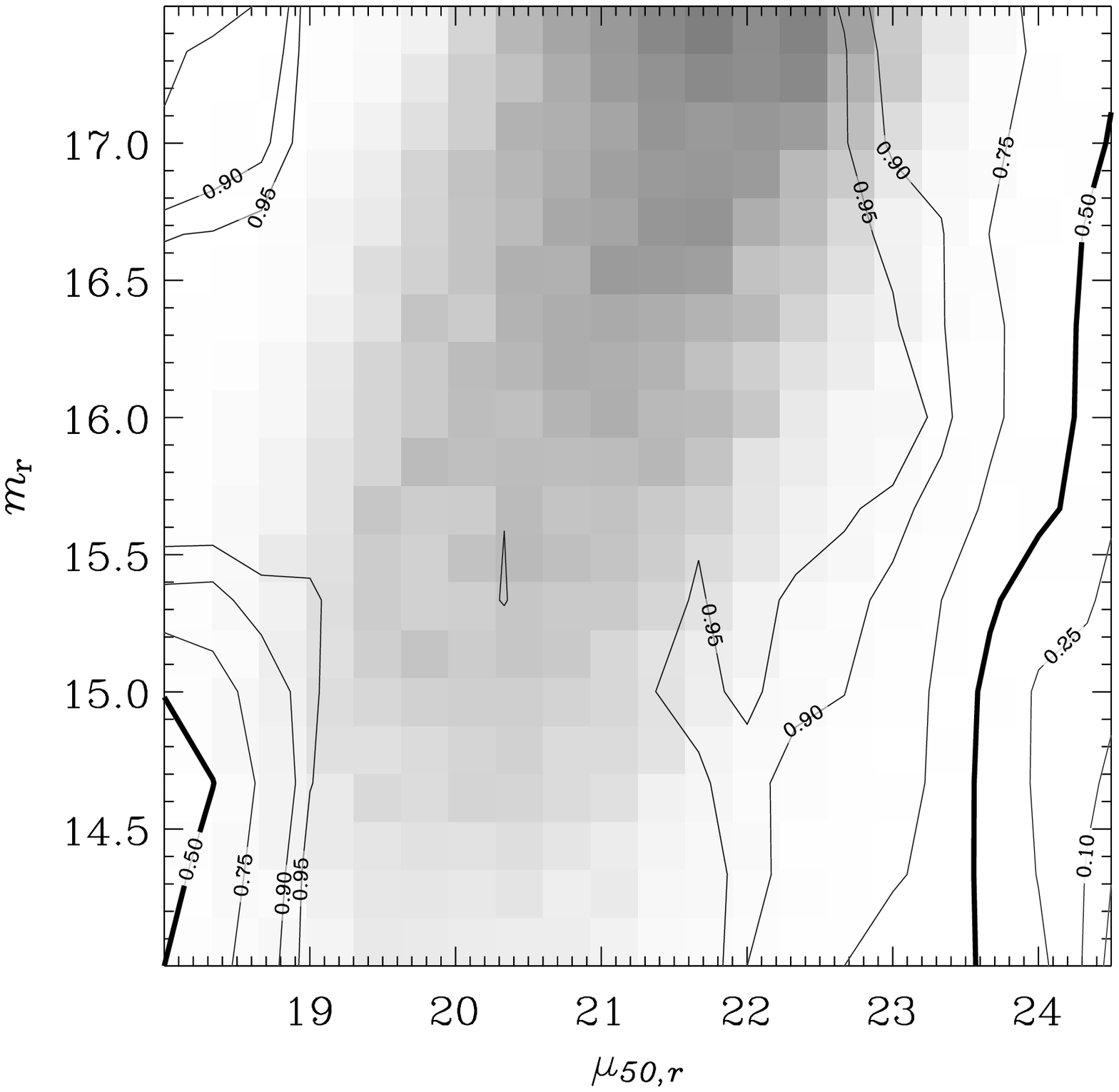}
\caption{\label{sbcomplete} Contours indicate the completeness of
  the SDSS photometric sample used here as a function of half-light
  surface brightness and magnitude, as determined from the fake data
  pipeline described in the text. The underlying greyscale image is
  the distribution of real galaxies in the low-redshift catalog. }
\end{figure}

The contours in Figure \ref{sbcomplete} show the completeness thus
calculated as a function of input half-light surface brightness and
input total magnitude. The underlying greyscale image is the
distribution of galaxies in the real galaxy catalog. Several features
are prominent. First, in the lower left, small, bright objects have a
low completeness. This occurs because of the lower limit on size of
$\theta_{50} > 2$ arcsec imposed in the Main sample selection criteria
for objects with $m_r < 15.5$ in order to eliminate double stars
(\citealt{strauss02a}). Second, in the upper left, there is
incompleteness in faint, high surface brightness objects. This
incompleteness results from the star-galaxy separator, which confuses
such objects for stars in marginal seeing (such as in the SDSS fields
we are testing here, for which the seeing is 1.5$''$ or so). We find
that in better seeing conditions, this corner of the plot is nearly
complete. Neither of these features are of particular interest here
(though it is worth noting that there is a class of compact galaxies
that this incompleteness affects, among them M32 and the UCDs of
\citealt{deady02a}). The greyscale shows that very few objects are
observed in these regimes; in addition, since low luminosity objects
tend to be lower surface brightness than average, these
incompletenesses at high surface brightness are unimportant for the
low luminosity galaxies.

However, on the right side it is clear that there is significant
incompleteness at low surface brightness (more or less independent of
size). A very small fraction of all SDSS galaxies are affected by this
incompleteness, since the galaxy distribution declines long before it
becomes important. However, low luminosity galaxies are
disproportionately affected by the incompleteness, because they tend
to be low surface brightness, and so for the purposes of this paper we
must consider this effect.

The incompleteness at low surface brightness is due primarily to two
effects. First, there is sometimes inappropriate shredding by the
photometric deblender of low surface brightness galaxies, often
related to the presence of nearby stars. This shredding tends to
reduce the galaxy fluxes to well below the flux limits of the
survey. Second, for many galaxies the flux is significantly reduced
because the sky subtraction determination (a 100 by 100 arcsec median
filter) subtracts a substantial fraction of the galaxy light
(\citealt{strauss02a}).

\begin{figure}
\figurenum{\fignum}
\plotone{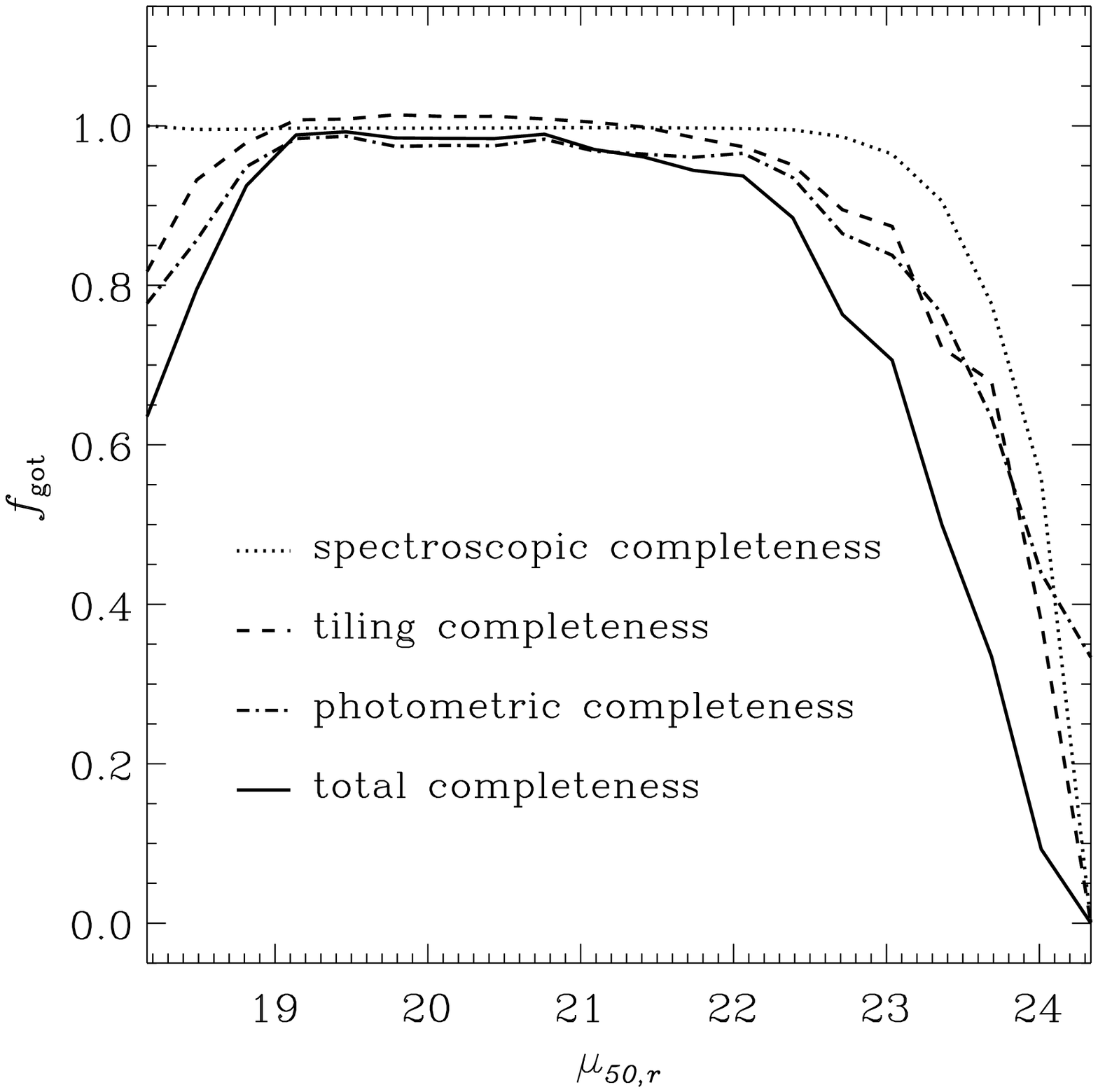}
\caption{\label{totalsb} Solid line is the completeness as a function
  of surface brightness (for a direction in which
  $f_{\mathrm{got}}(\alpha, \delta) = 1$. The contribution due to
  spectroscopic incompleteness is the dotted line. The contribution
  due to the tiling incompleteness is the dashed line. The
  contribution due to photometric incompleteness (estimated from
  simulations) is the dot-dashed line. }
\end{figure}

At low surface brightness the completeness is nearly independent of
apparent magnitude in Figure \ref{sbcomplete}. Therefore, we compress
the information by averaging across apparent magnitude, weighting by
the number of objects at each magnitude. This procedure yields the
dot-dashed curve in Figure \ref{totalsb}, the photometric catalog
completeness $f_{\mathrm{ph}}(\mu_{50,r})$ as a function of half-light
surface brightness.  Table \ref{totalsb_table} gives the values used
to make this figures.

We emphasize that this examination reveals the limits of the SDSS {\it
software}, not the SDSS {\it data}. The photometric pipeline was not
optimized for finding low-surface brightness objects, and there are
many more detectable objects in the data. For example, essentially
{\it every} object in our simulations (down to $\mu_{50,r} = 26$)
yields a large contiguous region of $5\sigma$ detected pixels; the
non-detections of the catalog are due to deblending and mis-estimates
of the sky. This accords with the results of \citet{cross04a}, who
find for the deeper Millenium Galaxy Catalog that $\sim 90\%$ of their
objects yielded SDSS detections even at $\mu_B \sim 26$.  We further
note that {\it all} the galaxies in the nearby catalog of
\citet{karachentsev04a} that fall within the SDSS area are of high
signal-to-noise ratio in the raw data and flagged as pixels with
signal, even if in the catalog they are deblended into many shreds or
otherwise excluded from the spectroscopic survey, as many of these
very nearby ($< 10$ Mpc) galaxies are. These facts motivate an effort
to recover even lower surface brightness objects from the data and to
weed out the considerable background of data defects, such as internal
reflections of bright stars. Such an effort has been begun by
\citet{kniazev04a}.

\subsection{Empirical test using Galactic extinction}
\label{dusttest}

\begin{figure}
\figurenum{\fignum}
\plotone{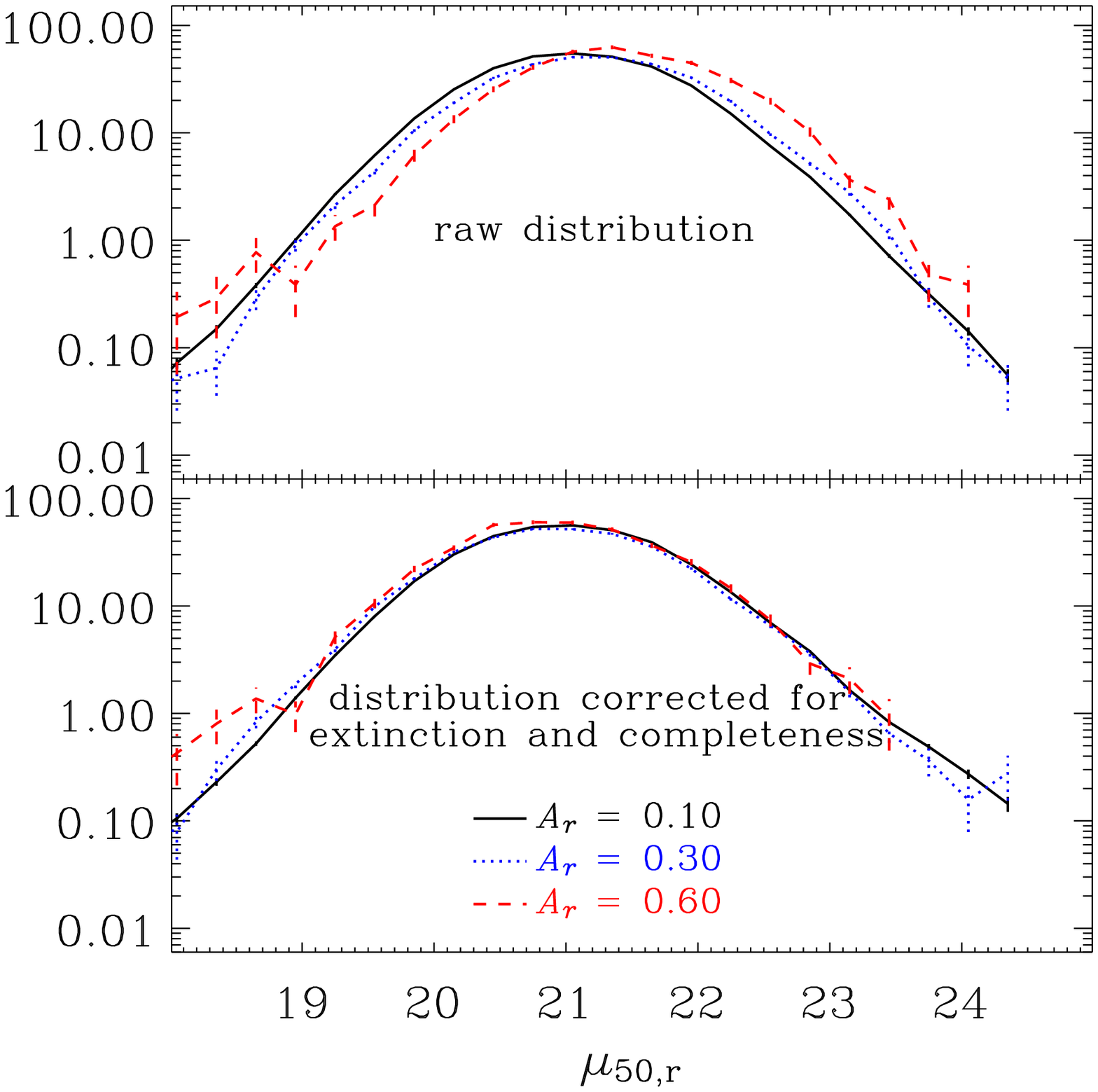}
\caption{\label{sb_vs_dust} Top panel shows the distribution of
  galaxies as a function of surface brightness (determined from the
  \Sersic\ fit) uncorrected for Galactic extinction, in three bins of
  extinction, as labeled. Bottom panel shows the same distribution
  corrected for Galactic extinction and surface brightness
  incompleteness. There is no evidence for additional incompleteness. }
\end{figure}

We can test the surface brightness completeness of the photometric
catalog in a different manner, following \citet{davies93a}. The
observed surface brightness distribution of galaxies is shifted
faintward by Galactic extinction by an amount that depends on
direction and is predicted by \citet{schlegel98a}. The top panel of
Figure \ref{sb_vs_dust} shows this trend for all galaxies with an
extinction-corrected Petrosian magnitude $m_r < 18$.  We use three
bins of Galactic extinction centered on the values shown. Each curve
is the surface density of galaxies as a function of Petrosian
half-light surface brightness for the given Galactic extinction, in
units that are the same for each bin of extinction.

The distribution of intrinsic surface brightness should not depend on
the amount of intervening Galactic dust. So if we correct the surface
brightness of each galaxy in the sample for the estimated dust
extinction, {\it and} correct the surface density using the
completeness estimate above (according to the observed surface
brightness, of course), the curves should be independent of the
extinction. Indeed, the bottom panel of Figure \ref{sb_vs_dust} shows
these distributions, which lie on each other quite well. 

These figures make clear two points. First, for the {\it bulk} of SDSS
galaxies, surface brightness completeness issues are
irrelevant. Second, we have reasonable estimates of our completeness
at the lowest surface brightness end, which we will use below to
attempt to judge what constraints we can make on the low luminosity
slope of the luminosity function.

\subsection{The tiling catalog} 
\label{ti_fgot}

The above tests evaluate the completeness of the catalog of SDSS
objects using the most recent version of the photometric
pipeline. However, as we found above, the completeness has as much to
do with the photometric software as it has to do with the data itself.
Furthermore, the software has changed over the course of the survey.
Of particular interest here, the deblender has improved substantially.
Earlier versions tended to incorrectly shred large galaxies into
multiple parts. When low surface brightness galaxies were shredded
they often had their flux reduced enough that no part of them was
targeted.  Although, as found above, this problem persists in the
later reductions, software improvements have greatly mitigated it. All
of the spectroscopic targeting used here was performed on the {\tt
v5\_2} version of the photometric pipeline, or previous.  Thus, we
need to evaluate what fraction of the objects detected in the {\tt
v5\_4} results actually were targeted by the old version of the
software.

To evaluate this completeness we ask what fraction of NYU-VAGC
galaxies brighter than the spectroscopic flux limit in each region of
sky (see \citealt{blanton04a}) have matches in the ``tiling catalog,''
the list of targets actually selected for spectroscopy by the SDSS.
We denote this fraction $f_{\mathrm{ti}}(\mu_{50,r})$.

There is a subtlety we must account for when considering the
completeness of the survey, which is that the tiling completeness
integrated over surface brightness is already included in the
completeness as a function of position.  The completeness as a
function of position $f_{\mathrm{got}}(\alpha,
\delta)$, given by the NYU-VAGC sample described in \citet{blanton04a},
is the fraction of galaxies brighter than the flux limit that are in
the spectroscopic catalog, which is basically a subset of the tiling
catalog.  Thus, $f_{\mathrm{got}}(\alpha, \delta)$ effectively
includes the tiling incompleteness integrated over all surface
brightnesses. So in order to isolate just the surface brightness
dependence of the tiling completeness, we scale our result according
to the total tiling completeness:
\begin{equation}
\tilde{f}_{\mathrm{ti}} = \frac{f_{\mathrm{ti}}(\mu_{50,r})}
{\int d\mu_{50,r} p(\mu_{50,r}) f_{\mathrm{ti}}(\mu_{50,r})}
\end{equation}
where $p(\mu_{50,r})$ is the probability density distribution of
surface brightness.  In practice, the denominator is around
0.97. Naturally, then, $\tilde{f}_{\mathrm{ti}}$ can exceed unity
(slightly), which is compensated for on average by the fact that
$f_{\mathrm{got}}(\alpha, \delta)$ is less than unity.

Figure \ref{totalsb} shows this completeness
$\tilde{f}_{\mathrm{ti}}(\mu_{50,r})$ as a function of surface
brightness as the dashed line, and Table \ref{totalsb_table} lists the
values plotted.

\subsection{The spectroscopic catalog}
\label{sp_fgot}

We determine the spectroscopic incompleteness as a function of surface
brightness simply by looking at what fraction of Main sample targets
whose spectra we attempted actually yielded reliable redshifts.
Figure \ref{totalsb} shows this fraction $f_{\mathrm{sp}}(\mu_{50,r})$
as a function of surface brightness; the bottom panel shows the number
of objects used in each bin.

\subsection{The total surface brightness completeness function}
\label{fgot}

In order to evaluate the total surface brightness completeness, we
simply multiply together the three components above.  Thus, the total
completeness as a function of position and surface brightness may be
expressed:
\begin{equation}
f(\mu_{50,r}, \alpha, \delta) = 
\tilde{f}_{\mathrm{ti}}(\mu_{50,r})
{f}_{\mathrm{sp}}(\mu_{50,r})
{f}_{\mathrm{ph}}(\mu_{50,r}) f_{\mathrm{got}}(\alpha, \delta)
\end{equation}
Figure \ref{totalsb} shows this function (for a direction in which
$f_{\mathrm{got}}(\alpha, \delta)=1$) as the solid line.

\section{Luminosity function and properties of galaxies}
\label{properties}

\subsection{Calculating the luminosity function}
\label{sbcorr}

Because low luminosity galaxies tend to be low surface brightness, the
surface brightness selection effects investigated in the previous
section are likely to cause an underestimate of the luminosity
function at low luminosities.  Below, we will present three versions
of the luminosity function:
\begin{enumerate}
\item the raw luminosity function for galaxies with surface
brightnesses $\mu_{50,r} < 24$, completely uncorrected for surface
brightness selection effects;
\item the luminosity function for
galaxies with surface brightnesses $\mu_{50,r} < 24$, corrected for
surface brightness incompleteness; and
\item an estimate of the total
luminosity function using a model for the luminosity-surface
brightness relationship.
\end{enumerate}
The first two we calculate in all five bands; the last we only perform
in the $r$ band.  

We calculate the uncorrected luminosity function (1) for galaxies with
$\mu_{50,r} < 24$ using the step-wise maximum likelihood method
(\citealt{efstathiou88a}).  An advantage of this method, compared to
the \Vmax\ method (\citealt{schmidt68a}) or the method of
\citet{blanton03c}, is that those methods are susceptible to cosmic
variance errors due to large-scale structure in the volume
probed. That is, overdensities or voids can lead to artificial
structure in the resulting luminosity function (or, in the case of the
method of \citealt{blanton03c}, artificial number density evolution).
Basically, the idea of
\citet{efstathiou88a} is to maximize the conditional likelihood of
observing each galaxy to have its luminosity {\it given} its
redshift. The exact method used here is described in more detail by
\citet{blanton01a}.

To calculate the luminosity function (2) for galaxies with $\mu_{50,r}
< 24$, we use these same techniques but include only galaxies in that
range of surface brightness and, in addition, weight each galaxy with
$1/f(\mu_{50,r})$, the inverse of the completeness found in the
previous section. In practice, this means taking the individual
likelihood for each galaxy to the {\it power} $1/f(\mu_{50,r})$.

Finally, to estimate the numbers of galaxies missing below $\mu_{50,r}
=24$ and thus the ``total'' luminosity function (3), we need to assume
some model for the relationship between surface brightness and
luminosity. To do so, we select galaxies with $n<2$ and $M_r <
-18$. For such galaxies, the surface brightness limits are
unimportant, as we will see below.  We fit the following model to
their conditional distribution in surface brightness as a function of
magnitude:
\begin{equation}
\label{purecheese}
P(\mu_{50,r} | M_r) = \frac{1}{\sqrt{2 \pi} \sigma_\mu} \exp
\left[(\mu_{50,r} - {\bar\mu_{50,r}}(M_r))^2 / 2 \sigma_\mu^2(M_r) \right]
\end{equation}
where
\begin{eqnarray}
{\bar\mu_{50,r}}(M_r) &=& \mu_{50,\ast} + \gamma (M_r +20.5) \cr
\sigma_\mu(M_r) &=& \sigma_{\mu\ast} + \beta (M_r + 20.5) 
\end{eqnarray}
The distribution is thus a Gaussian in surface brightness at all
absolute magnitudes, with a power law relationship between absolute
magnitude and the mean (linearly expressed) surface brightness. We
allow the width of the distribution to grow linearly as the absolute
magnitude increases.  Note that we do not specify the normalization at
each absolute magnitude. In this definition, $\mu_{50, \ast}$ is the
typical surface brightness of an exponential galaxy around $M_\ast
\sim -20.5$. Our best fit is $\gamma = 0.45$, $\mu_{50, \ast} =
20.56$, $\beta=0.081$, and $\sigma_{\mu\ast} = 0.58$.

\begin{figure}
\figurenum{\fignum}
\plotone{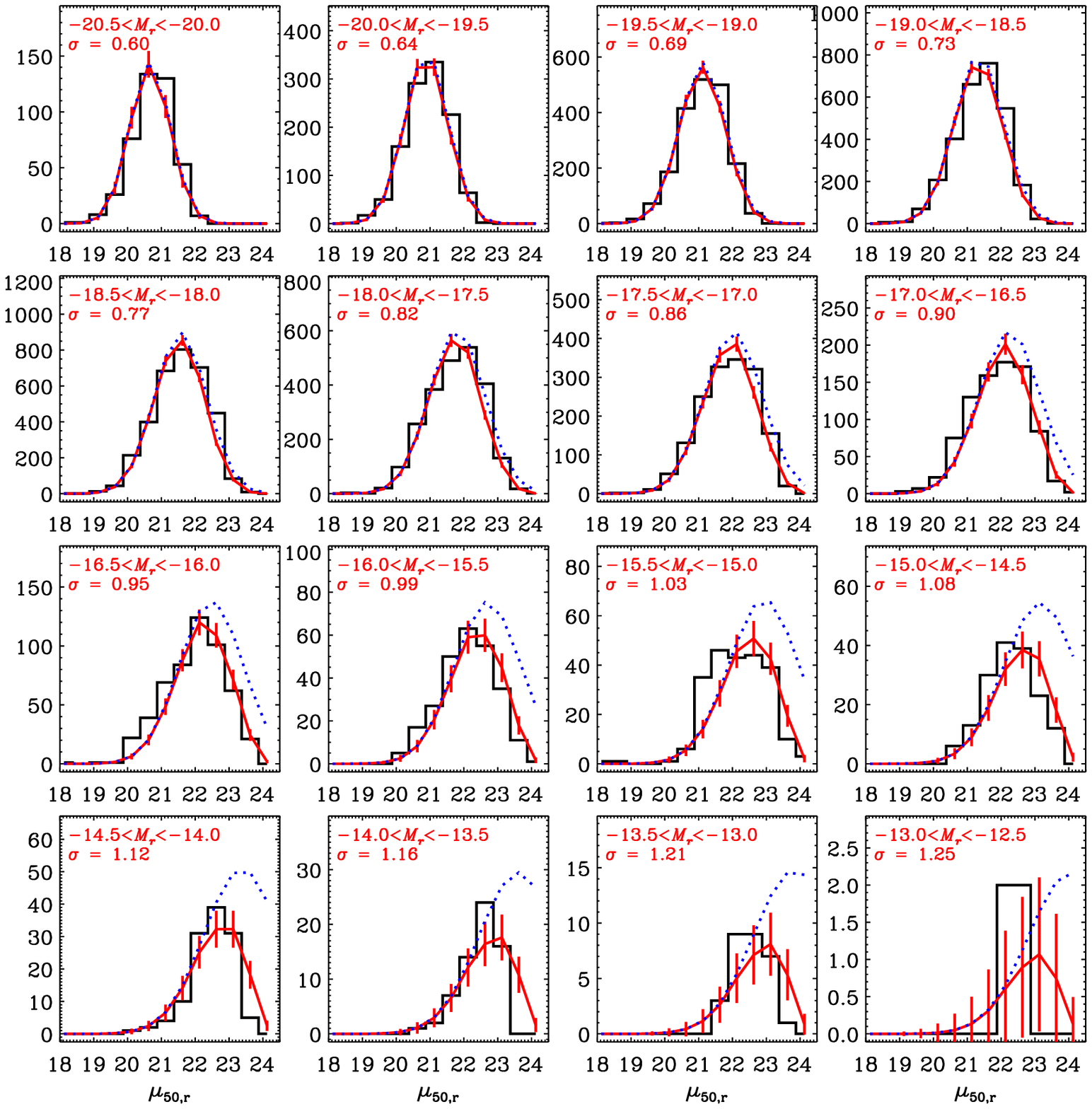}
\caption{\label{fit_absm_sb} Each panel shows as a histogram the
  distribution of surface brightness $\mu_{50,r}$ for a bin of
  absolute magnitude $M_r$. The dotted line is the model of Equation
  \ref{purecheese} as fit to galaxies more luminous than $M_r =
  -18$. The solid line is the model multiplied by the surface
  brightness completeness. }
\end{figure}

Figure \ref{fit_absm_sb} shows the surface brightness distribution in
bins of absolute magnitude. The histogram is the observed
distribution. The dotted line is the model of Equation
\ref{purecheese}. The solid line is the model multiplied by the
photometric and spectroscopic completenesses from Table
\ref{totalsb_table}. The lines are normalized to minimize the residuals
between the solid line and the solid histogram. The error bars on the
solid lines assume Poisson statistics.

The model is good in the range over which we have fit ($M_r <
-18$). As stated above, it is clear that in this regime galaxies lie
safely away from the surface brightness limits of the survey. Less
luminous than that ($M_r > -18$), for which the model represents an
extrapolation, there are some deviations. In particular, the observed
distribution appears to have a slightly larger tail on the high
surface brightness end than the model assumes.

\begin{figure}
\figurenum{\fignum}
\plotone{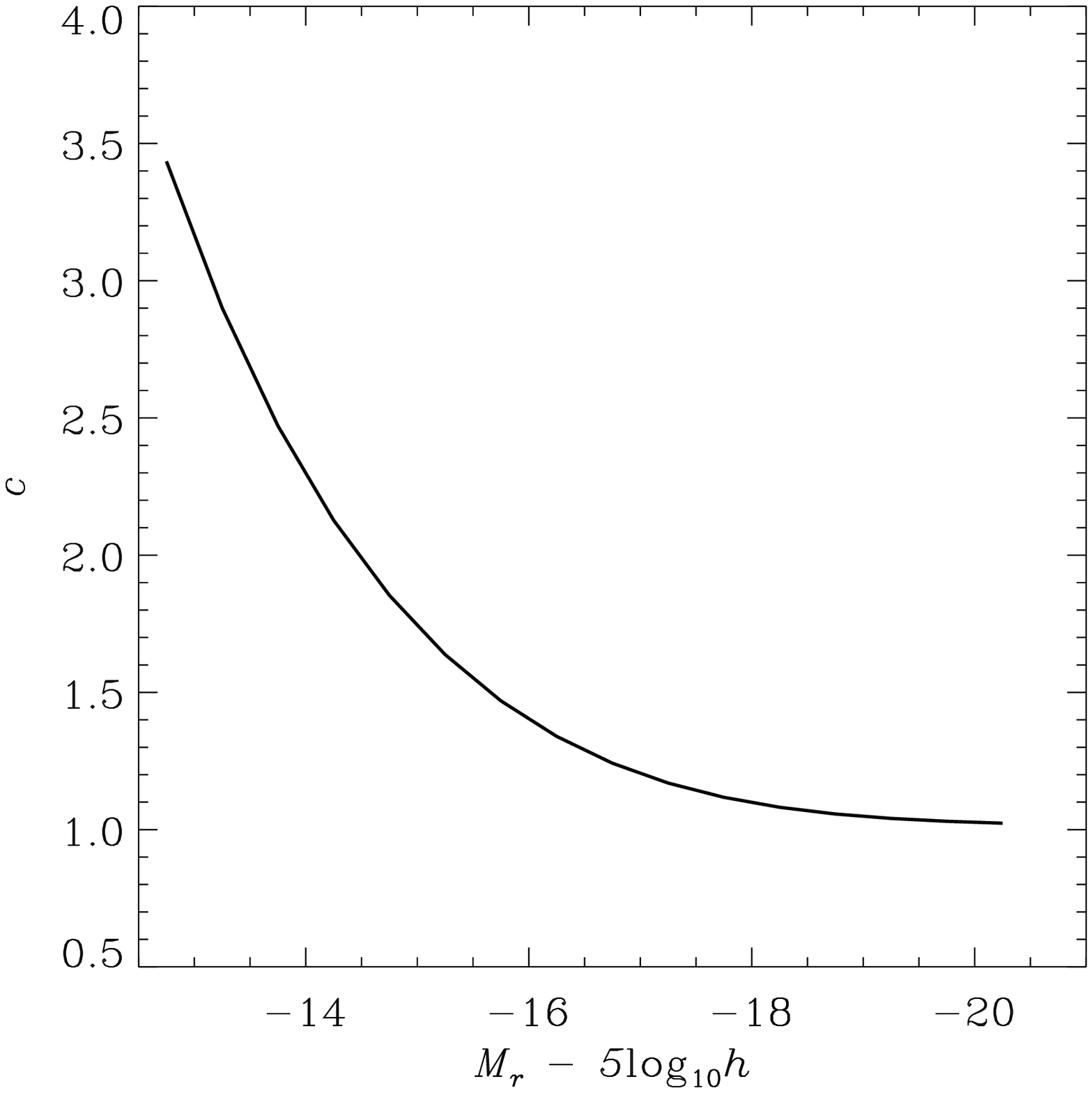}
\caption{\label{cfact} Estimate of possible correction factor due to
  surface brightness incompleteness in the SDSS, as a function of
  absolute magnitude. Based on fit of Equation \ref{purecheese} to
  luminosity surface brightness relationship at high luminosity ($M_r
  < -18$).}
\end{figure}

We use this model to estimate the possible magnitude of surface
brightness selection effects as $c$, the ratio of two integrals: that
of the solid line to that of the dotted line. That is, our estimate is
the fraction of galaxies in our model which would be observable given
our estimated completeness. Figure \ref{cfact} shows the resulting
``correction'' factor one should therefore apply as a function of
absolute magnitude. 

By weighting by $c$, we obtain the ``total'' luminosity function (3).
This estimate is obviously highly uncertain, since the model of
Equation \ref{purecheese} does not necessarily represent the surface
brightness distribution below our detection limits. However, it is our
best guess at what could reasonably be missing.

\subsection{The luminosity function and the luminosity density}

\begin{figure}
\figurenum{\fignum}
\plotone{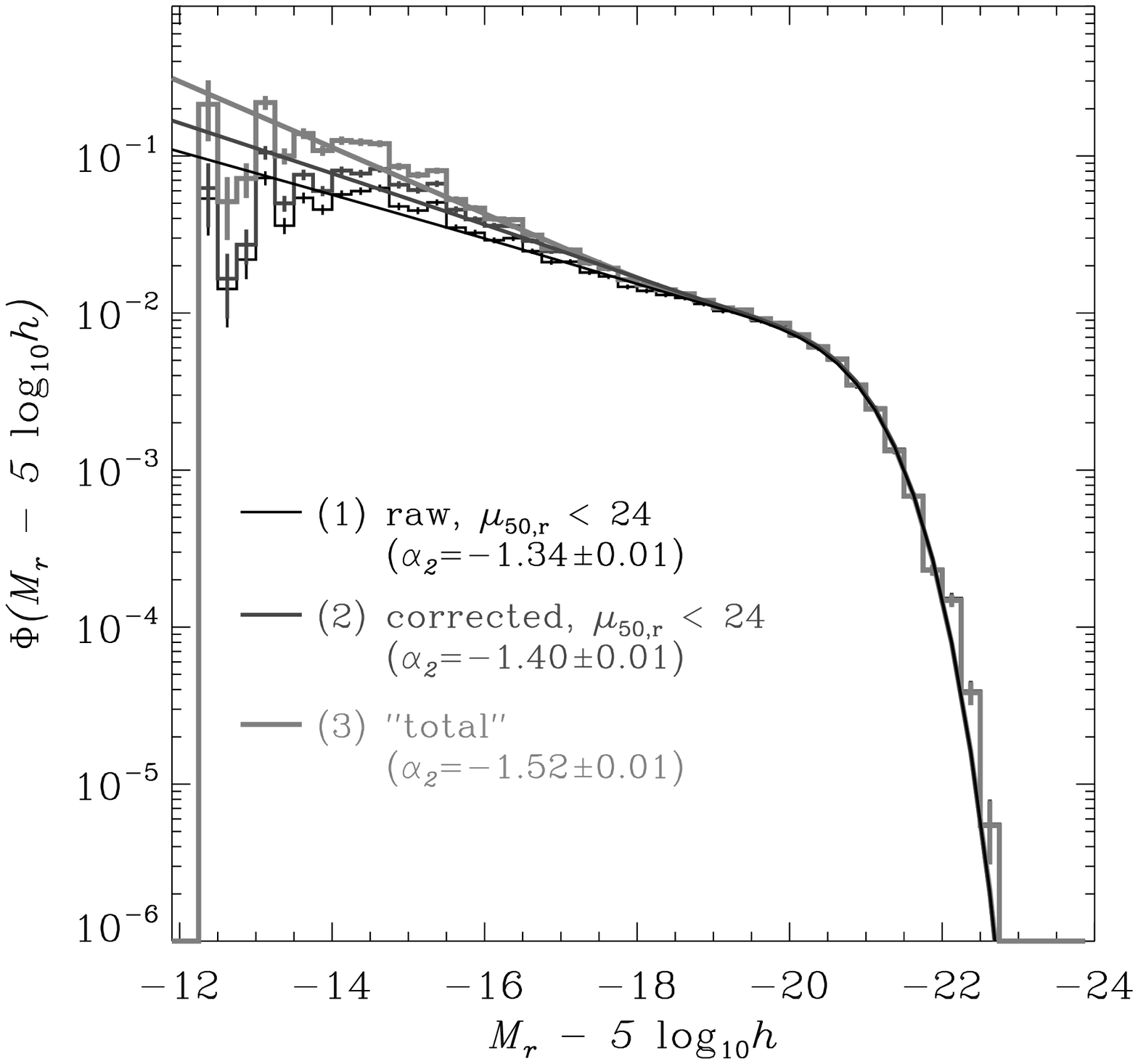}
\caption{\label{lf_cfact} The luminosity function in the
  $r$-band calculated using the step-wise maximum likelihood method,
  with bins of width 0.25 mag. The black histogram is the minimal
  luminosity function (1) for galaxies with $\mu_{50,r} < 24$,
  described in Section \ref{sbcorr}, with no correction for surface
  brightness selection effects. The dark grey histogram is the
  luminosity function for galaxies with $\mu_{50,r} < 24$ corrected
  for surface brightness incompleteness.  The light grey histogram is
  an attempt to estimate how many galaxies there might be using a
  simple model for the luminosity surface brightness relationship. The
  values used in this plot are given in Table \ref{lf_cfact}. The
  smooth curves are double Schechter function fits to each result,
  whose parameters are given in Table \ref{lf_cfact_sch}. The bottom
  panel is the number of galaxies contributing to each bin. All
  magnitudes here and elsewhere in the paper are $K$-corrected to
  rest-frame bandpasses and have no evolution correction applied.}
\end{figure}

Figure \ref{lf_cfact} shows the luminosity function of our sample
calculated using the step-wise maximum likelihood estimator. The black
histogram shows (1), the luminosity function for galaxies with
$\mu_{50,r} < 24$ uncorrected for surface brightness incompleteness
and represents a ``minimum'' abundance of galaxies. The dark grey
represents (2), the luminosity function of galaxies with $\mu_{50,r} <
24$ corrected for incompleteness.  The light grey histogram represents
(3), the ``total'' luminosity function using the method of Section
\ref{sbcorr} to ``correct'' for missing galaxies below the surface
brightness limits. Note that this third estimate, while it represents
a reasonable extrapolation of the abundance of galaxies at these
luminosities, is not an actual measurement of it. We claim that an
abundance of low luminosity galaxies similar to the light grey
histogram cannot be ruled out from any current data. Table
\ref{lf_cfact_table} gives the values used for the histograms in
Figure \ref{lf_cfact}.

The smooth lines in Figure \ref{lf_cfact} represent fits to the
luminosity function using a double Schechter function:
\begin{equation}
\Phi(L) dL = \frac{dL}{L_\ast}
  \exp(-L/L_{\ast}) \left[
\phi_{\ast,1} 
\left( \frac{L}{L_{\ast}} \right)^{\alpha_1} + 
\phi_{\ast,2} 
\left( \frac{L}{L_{\ast}} \right)^{\alpha_2} 
  \right]
\end{equation}
Stated in terms of absolute magnitude $M = -2.5 \log_{10} (L) +$ const
this equation is:
\begin{equation}
\Phi(M) =  0.4 \ln 10 dM
  \exp\left(-10^{-0.4(M-M_{\ast})}\right) \left[
\phi_{\ast,1} 
10^{-0.4 \left( M-M_{\ast} \right)(\alpha_1+1)} + 
\phi_{\ast,2} 
10^{-0.4 \left( M-M_{\ast} \right)(\alpha_2+1)}
  \right]
\end{equation}
This function is a much better fit than a single Schechter function,
which fails to capture the upturn in the luminosity function at $M_r
\sim -18$ or so. Although the luminosity function is nearly flat at
luminosities higher than that (and less than $L_{\ast}$), even the
uncorrected, ``minimal'' luminosity function turns up to a slope of
about $\alpha_2 \sim -1.3$ below that luminosity. As shown in the fit
to the solid histogram, surface brightness selection effects may imply
a much steeper slope, possibly steeper than $\alpha_2 \sim -1.5$. The
full set of parameters found in these fits are listed in Table
\ref{lf_cfact_sch} (listed in terms of $M_\ast$).

\begin{figure}
\figurenum{\fignum}
\plotone{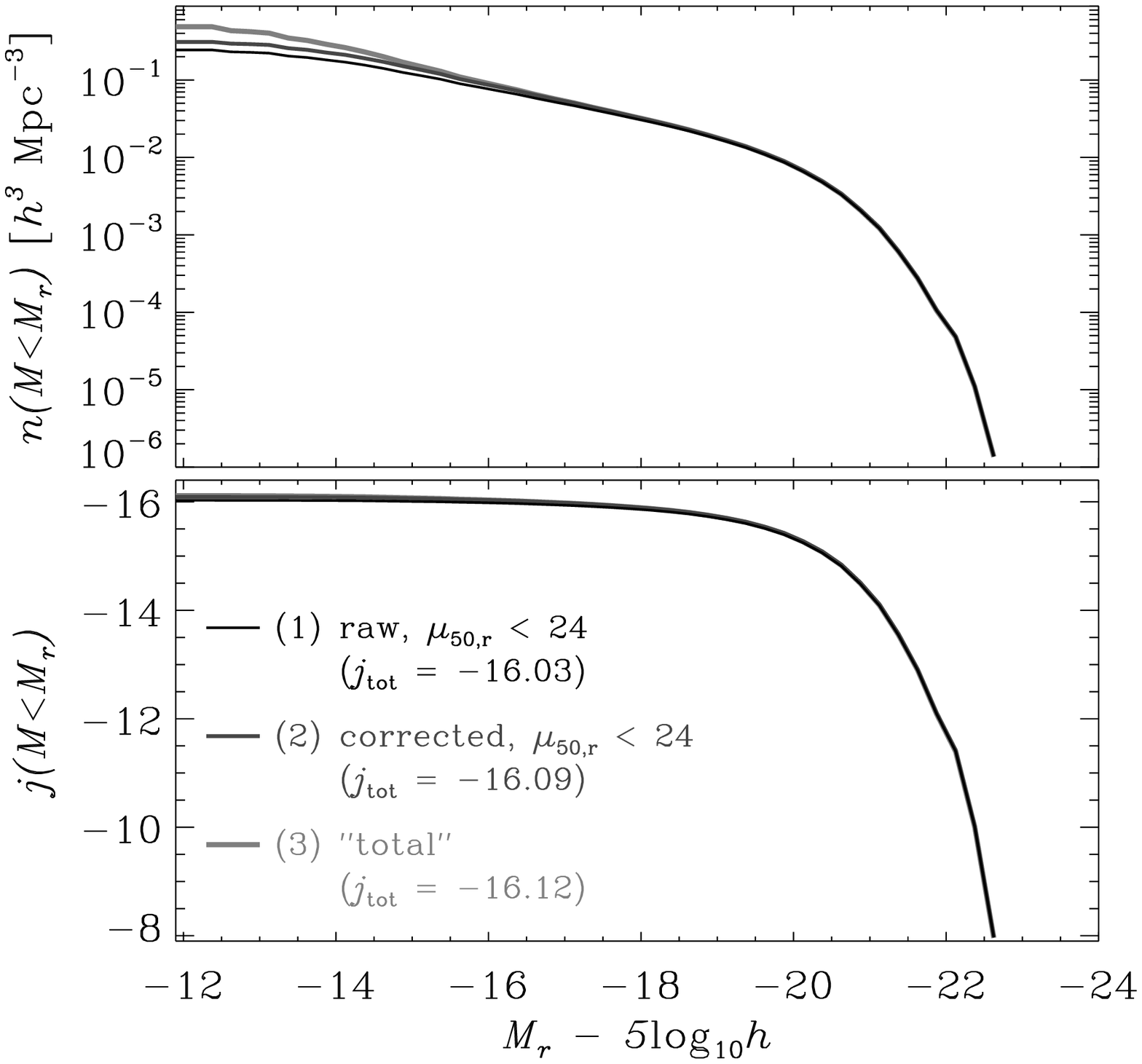}
\caption{\label{mden} {\it Top panel:} Total number density in
galaxies more luminous than $M_r$, as a function of $M_r$, for all
three luminosity functions of Figure \ref{lf_cfact}. The black line is
the raw luminosity function for galaxies with $\mu_{50,r} < 24$ (1);
the dark grey line is the corrected luminosity function for galaxies
with $\mu_{50,r} < 24$ (2), and the dashed line is our estimate of the
``total'' luminosity function (3), all as described in Section
\ref{sbcorr}.  {\it Bottom panel:} Same as top panel for total 
luminosity density. In all cases, most of the luminosity density
($\sim 90\%$) is contained in galaxies with $M_r< -17$. }
\end{figure}

The top panel of Figure \ref{mden} shows the cumulative number density
distribution of galaxies brighter than $M_r$ as a function of $M_r$,
for all three versions of the luminosity function.  The bottom panel
of Figure \ref{mden} similarly shows the luminosity density, in units
of absolute magnitudes per $h^{-1}$ Mpc$^3$.  The total luminosity
density in these units is $-16.02$ for the uncorrected case (1), and
$-16.12$ for the ``total'' case (3).  Note that, even for the
``total'' case, most of the luminosity density ($> 90\%$) is contained
in galaxies with $M_r < -17$. That is, as noted by a number of authors
in the past (\citealt{mcgaugh96a, sprayberry97a, blanton01a, cross01a,
blanton03d}) the contribution of low luminosity and low surface
brightness galaxies to the overall optical luminosity density appears
small.

\begin{figure}
\figurenum{\fignum}
\plotone{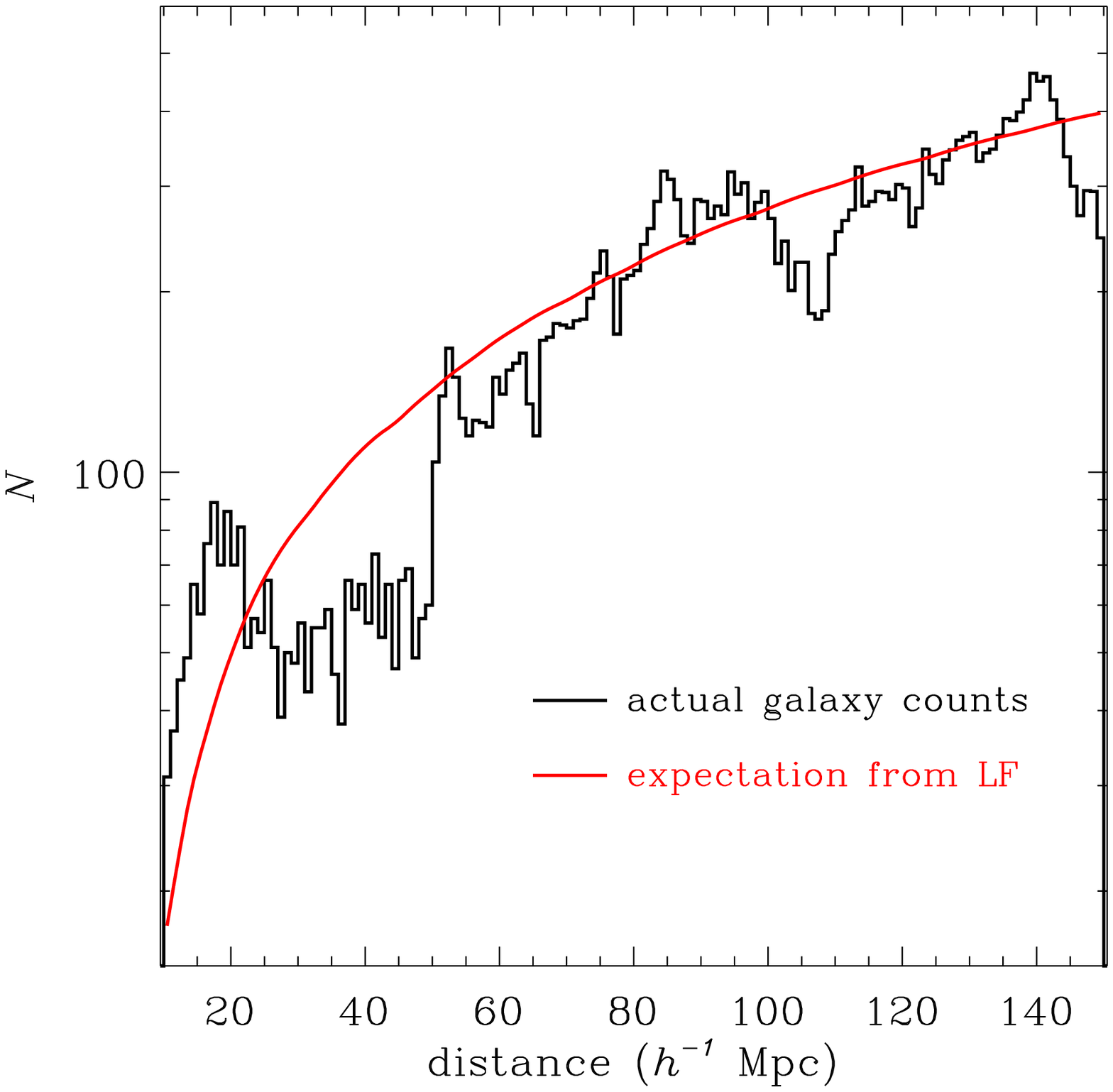}
\caption{\label{distance} Histogram of distances to galaxies in the
sample, in 1 $h^{-1}$ Mpc thick shells. The smooth line represents the
expected number in each bin based on the best fit raw luminosity
function (version 1) described in Section \ref{sbcorr}. }
\end{figure}

Figure \ref{distance} shows the distance distribution of galaxies
based on our best-fit distances. The smooth line is the expected
distribution given the raw luminosity function (1). Note there is
change by a factor of $\sim 2$ in the mean density of galaxies in the
local volume covered by this sample, which appears to be an
underdensity in the nearby section of the volume covered by this
sample. The fluctuation is within a volume of about 2 $\times$
$10^{4}$ $h^{-3}$ Mpc$^3$, equivalent to about a $\sim 20$ $h^{-1}$
Mpc radius sphere. Given the results of
\citet{tegmark04a}, whose constraints come from galaxy clustering over
much larger volumes, we expect $\sigma_{20} \sim 0.5$. Thus, the
structure we observe is consistent with what we know about clustering
of galaxies on large scales using more distant samples (which is
itself consistent with predictions based on cosmic microwave
background and other observations).

\begin{figure}
\figurenum{\fignum}
\plotone{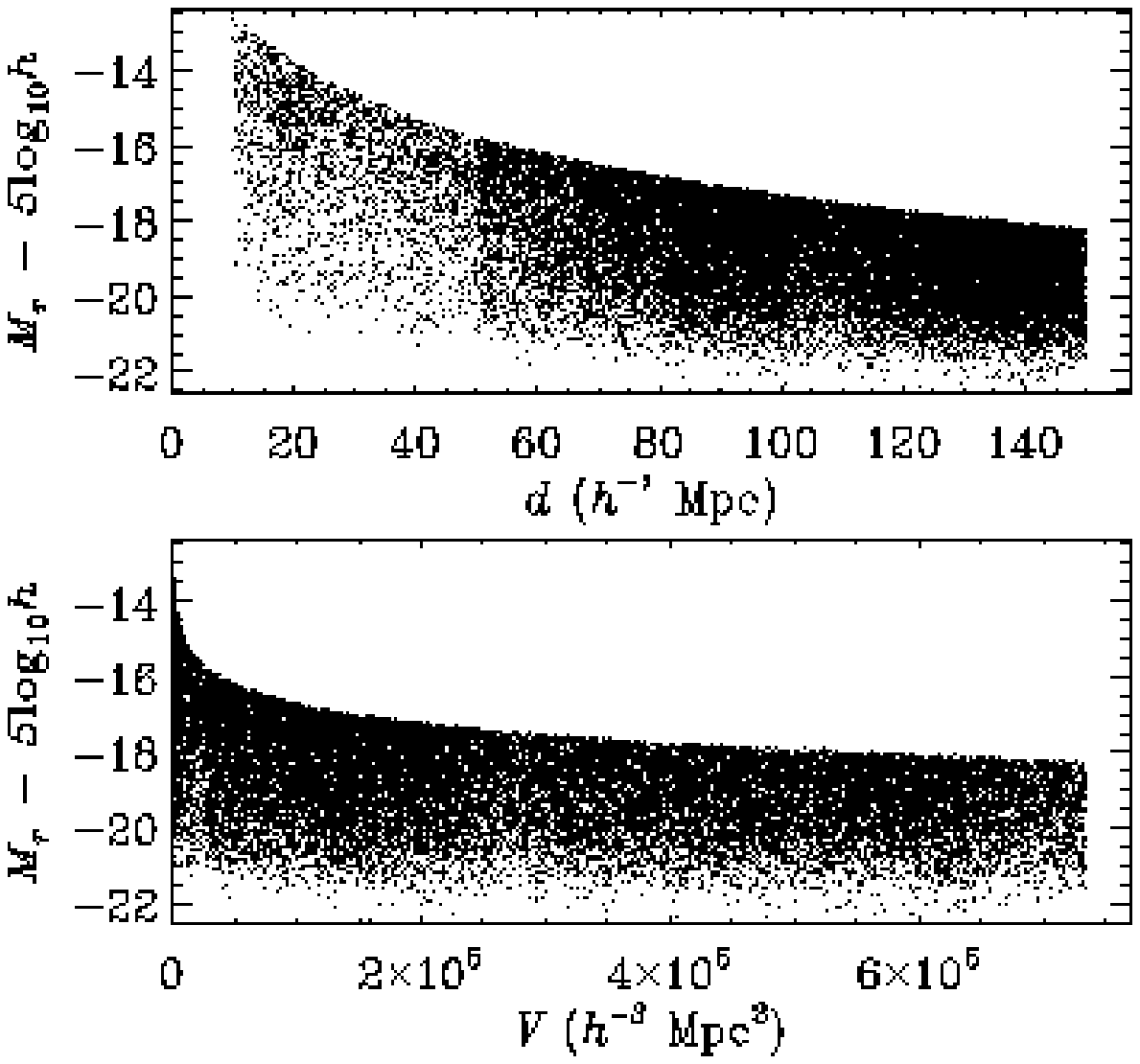}
\caption{\label{volume} Top panel shows the joint distribution of
distance and absolute magnitude.  Bottom panel shows the joint
distribution of enclosed volume and absolute magnitude. The
distribution of galaxies in this plot shows a fairly constant
distribution further than $z\sim 0.015$ ($V \sim 2\times 10^4$).
However, there is noticeable large-scale structure interior to that,
as well as some likely incompleteness at the luminous end of the
distribution.}
\end{figure}

Figure \ref{volume} shows the joint distribution of redshift and
absolute magnitude in the top panel, and of enclosed volume and
absolute magnitude in the bottom panel. Note there is a decrease in
the density of galaxies at small volumes, corresponding to the change
in the overdensity of galaxies seen in Figure \ref{distance}. For the
most luminous galaxies, there is something of a decline towards small
volumes, probably indicative of a bias against large objects due to
the SDSS selection process. This bias comes from the fact that large
galaxies are often deblended incorrectly by the SDSS photometric
pipeline, because they often have complex structure and because they
often run into the edges of fields. For future versions of this
catalog, we will concentrate on a more complete sample at the bright
end to resolve these problems; however, the focus of this paper is on
the low luminosity galaxies, so we ignore this potential problem for
the moment.

\begin{figure}
\figurenum{\fignum}
\plotone{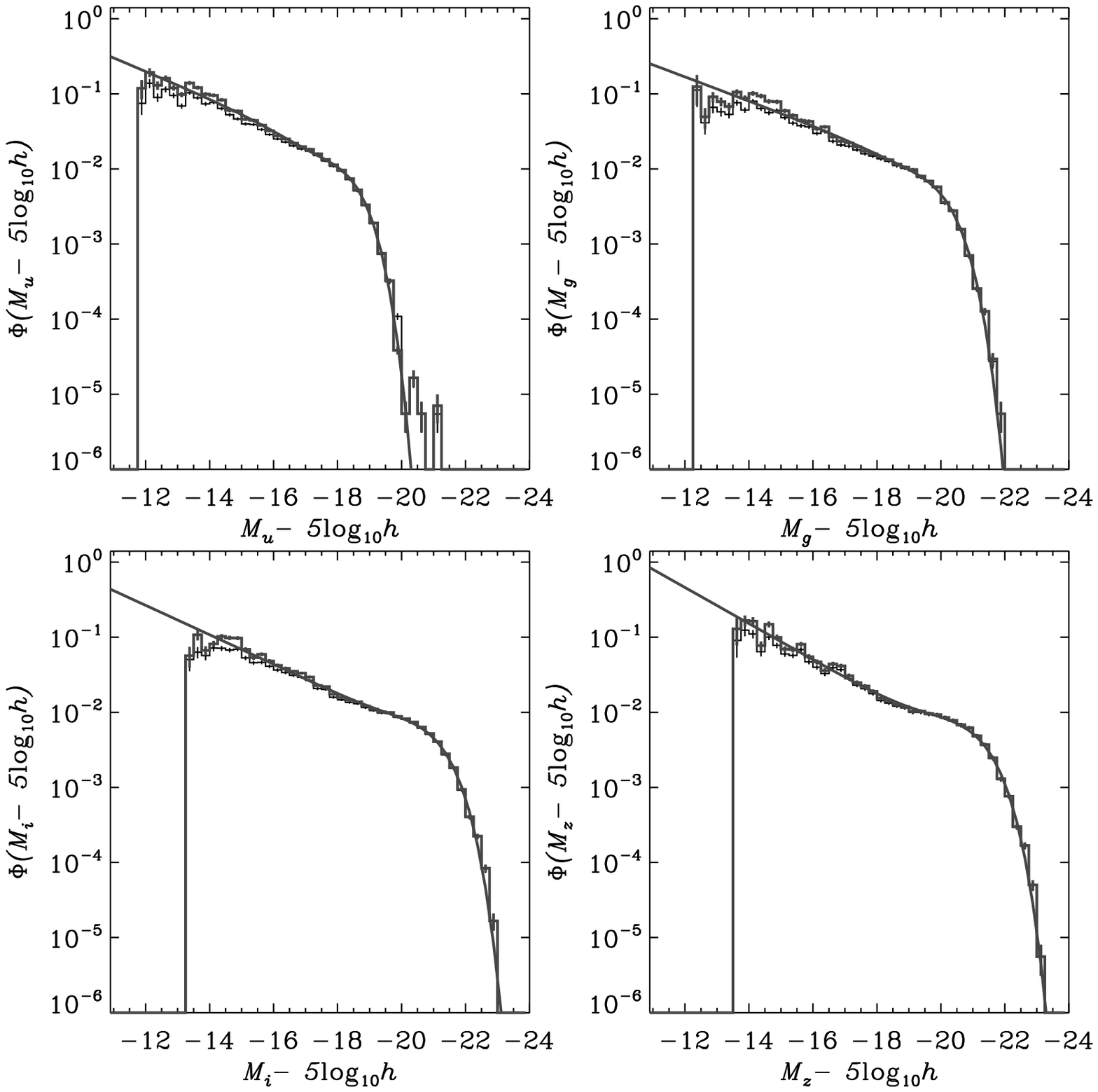}
\caption{\label{lf_bands} The histograms in each panel are similar to
that in the top panel of Figure \ref{lf_cfact}, for the $ugiz$ bands,
as labeled. The black histograms correspond to the raw luminosity
function for galaxies with $\mu_{50,r} < 24$, version (1) in Section
\ref{sbcorr}. The grey histograms correspond to the luminosity
function for galaxies with with $\mu_{50,r} < 24$ corrected for
surface brightness incompleteness, version (2) in Section
\ref{sbcorr}. For the $ugiz$ bands, we have not created a luminosity
function corresponding to version (3) in that section. The smooth
lines correspond to the double Schechter function fits the surface
brightness corrected luminosity function version (2), whose parameters
we list in Table \ref{lf_cfact_sch}. }
\end{figure}

In addition, we calculate the luminosity function in the $ugiz$ bands
using the step-wise maximum likelihood method and compare our result
to those of \citet{blanton03c}. In order to calculate each luminosity
function, we restrict our sample to galaxies for which $m_u < 18.4$,
$m_g < 17.7$, $m_i < 17.5$, or $m_z < 16.9$, depending on which band
we are interested in. At each magnitude limit we are highly complete
in the given band (\citealt{blanton01a}). We follow the same procedure
as we did in calculating the $r$-band luminosity function.  Figure
\ref{lf_bands} shows the results of this procedure for the raw
luminosity function for galaxies with $\mu_{50,r} < 24$ (1) as the
black histograms. The version for galaxies with $\mu_{50,r} < 24$
corrected for incompleteness (2) appears as the grey histograms. The
smooth lines are the double Schechter function fits we describe above
fit to each of the corrected luminosity functions. Tables
\ref{lf_ugiz_table} and \ref{lf_ugiz_2_table} tabulate the results for
the uncorrected and the corrected case, respectively. Table
\ref{lf_cfact_sch} lists the double Schechter function parameters for
each luminosity function. Note that \citet{baldry05a}
suspect large sky subtraction errors to add considerable scatter to
the $u$-band magnitudes of spectroscopic galaxies; our luminosity
function here does not use the empirical corrections they recommend.

\subsection{Comparison to the \Vmax\ method}

\begin{figure}
\figurenum{\fignum}
\plotone{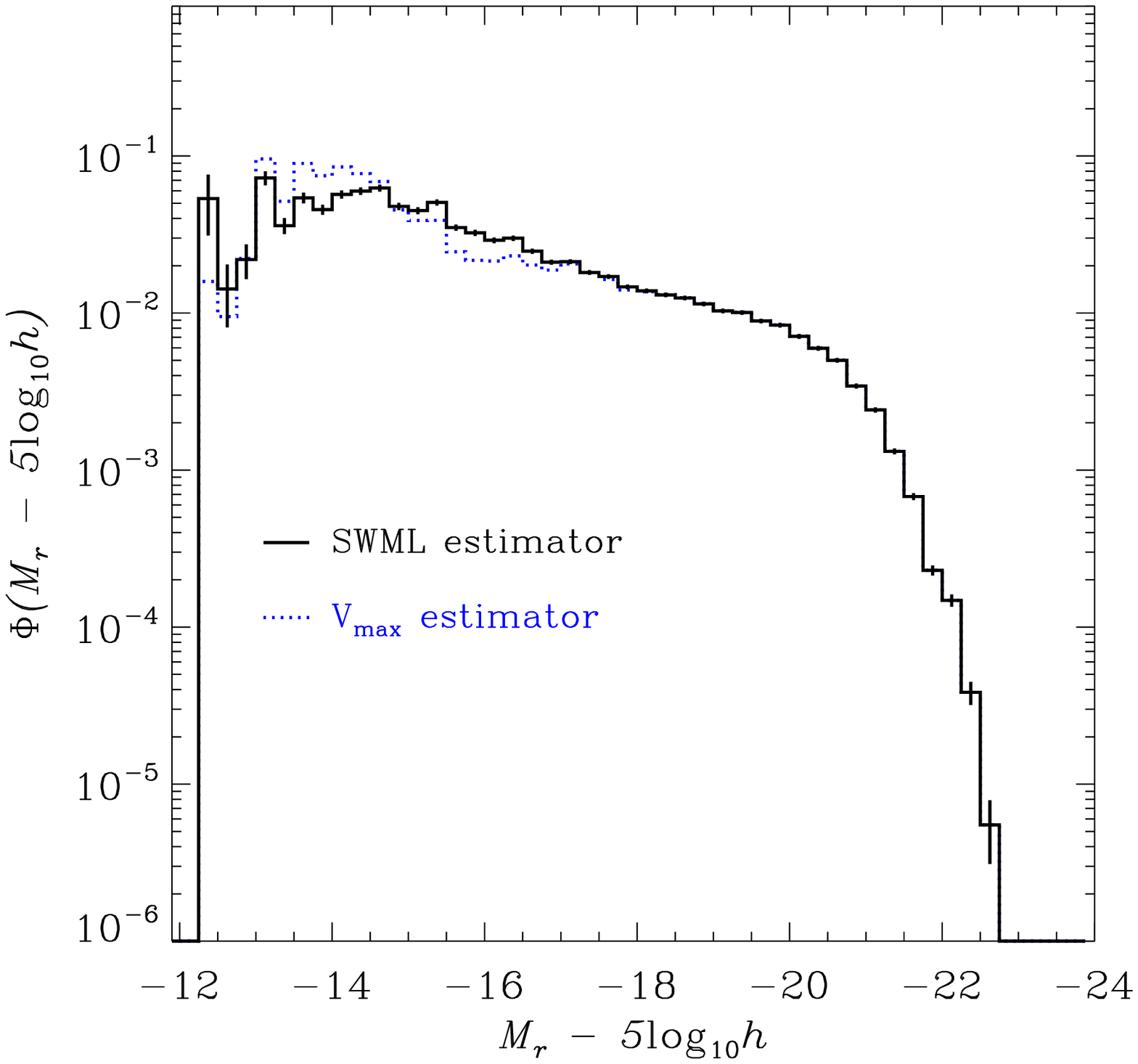}
\caption{\label{lf_vmax} Similar to Figure \ref{lf_cfact}, but showing
only the uncorrected results for the step-wise maximum likelihood
method (solid histogram) and the $\Vmax$ method (dotted
histogram). There is complete agreement at high luminosity, but
disagreement at low luminosity, probably due to large-scale structure
affecting the $\Vmax$ estimate. }
\end{figure}

Figure \ref{lf_vmax} shows a comparison of our original (uncorrected)
luminosity function using the step-wise maximum likelihood method and
that obtained using the $1/\Vmax$ method.  The $1/\Vmax$ method is
more susceptible to the effects of large-scale structure (dotted
histogram). These two estimates are in good agreement at the bright
end (where they probe a large amount of volume), though some of the
features in the $\Vmax$ result appear to be due to large-scale
structure. In particular, the overdensity closer than 20 $h^{-1}$ Mpc
and the underdensity in the range 30--50 $h^{-1}$ Mpc in Figure
\ref{distance} appear to result in a corresponding overestimate of the
luminosity function at about $M_r - 5 \log_{10} h \sim -14$ and an
underestimate at about $M_r - 5 \log_{10} h \sim -16$.  We show these
results to give a sense of the possible errors in the distributions of
galaxy properties weighted by $1/\Vmax$ shown in Section
\ref{props}. These errors are significant but not overwhelming. The
\Vmax\ results are listed in Table
\ref{lf_cfact_table}. 

\subsection{Comparison to the literature}

There exist previous SDSS results using the full sample of galaxies
out to a redshift $z=0.22$ (\citealt{blanton03c}). Those results are
quoted in the $\band{0.1}{ugriz}$ bands, which are the $ugriz$ bands
shifted blueward by a factor 1.1 (\citealt{blanton03b}), and are
evolution correction to an effective redshift of $z=0.1$.  We can
calculate the photometric transformations from $\band{0.1}{ugriz}$ to
$ugriz$ based on the SED models fit by {\tt kcorrect v3\_2}, for
galaxies of median intrinsic $\band{0.1}{(g-r)}$ color. This procedure
yields the relationships:
\begin{eqnarray}
  u &=& \band{0.1}{u} -0.38 \cr
  g &=& \band{0.1}{g} -0.41 \cr
  r &=& \band{0.1}{r} -0.22 \cr
  i &=& \band{0.1}{i} -0.19 \cr
  z &=& \band{0.1}{z} -0.11 
\end{eqnarray}
We correct for evolution using the
luminosity evolution model of \citet{blanton03c}:
\begin{eqnarray}
  M(z) &=& M (z = 0.1) - (z-0.1) Q  \cr
  {\bar n}(z) &=& {\bar n}( z=0.1) 10^{0.4 (z-0.1) P}
\end{eqnarray}
where $Q= 4.22, 2.04, 1.62, 1.61, 0.76$ and $P= 0.20, 0.32, 0.18,
0.58, 2.28$ in the $ugriz$ bands. 

\begin{figure}
\figurenum{\fignum}
\plotone{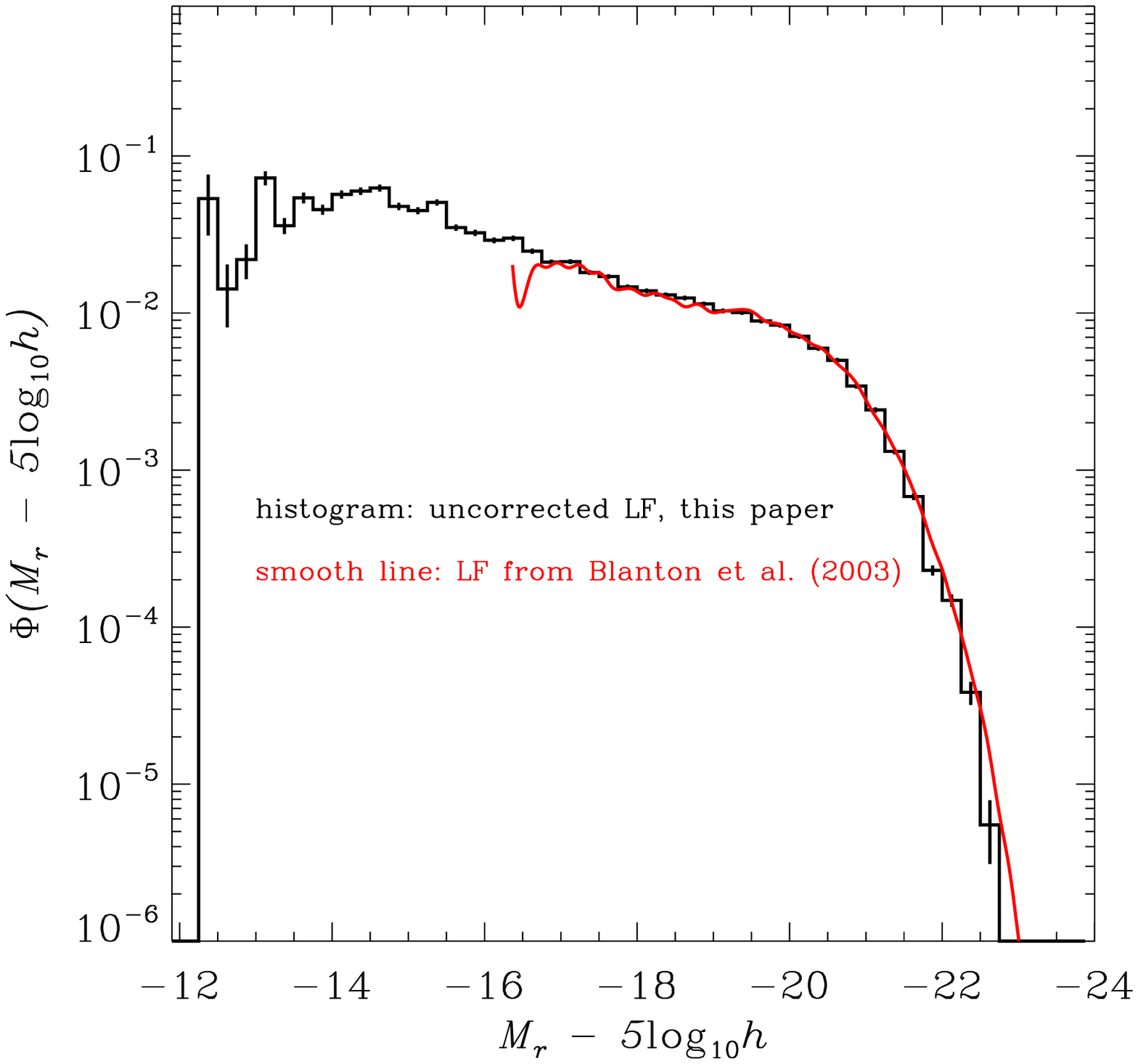}
\caption{\label{lf_old} Similar to Figure \ref{lf_cfact}, but showing
only the uncorrected results for the step-wise maximum likelihood
method (histogram) compared to the results of \citet{blanton03c}. 
The two methods agree well. }
\end{figure}

Figure \ref{lf_old} shows the raw luminosity function (1) in the
$r$-band for our sample compared to the results of \citet{blanton03c},
shown as the solid line.  Figure \ref{lf_bands} shows the comparison
for the $ugiz$ bands.  The agreement between the determinations is
good in the range measured by \citet{blanton03c}.

It is worth noting that both the current results and those of
\citet{blanton03c} differ markedly from those of
\citet{blanton01a}, whose results are incorrect because they ignored
the effects of evolution on the normalization and slope of the
luminosity function. These effects are fully described in
\citet{blanton03c}, which describes exactly why the \citet{blanton01a}
is in error. Essentially, the volume-weighted normalization allowed
the tiny number density of high luminosity objects to carry large
statistical weight. Since the luminosity function is so steep at high
luminosity, a small evolution in absolute magnitude resulted in a
large change in number density at fixed absolute magnitude. For this
paper, we have been able to ignore the effects of evolution because we
are considering such a small redshift range. For example, the $Q=1.62$
evolution in the $r$-band found by \citet{blanton03c} leads to a
magnitude difference across this sample of only about $0.08$. In
addition, this sample is volume-limited for $M_r < -18.5$, so the
steepness of the luminosity function at the luminous end is not as
important as it was in the more distant sample of
\citet{blanton01a}. 

\begin{figure}
\figurenum{\fignum}
\plotone{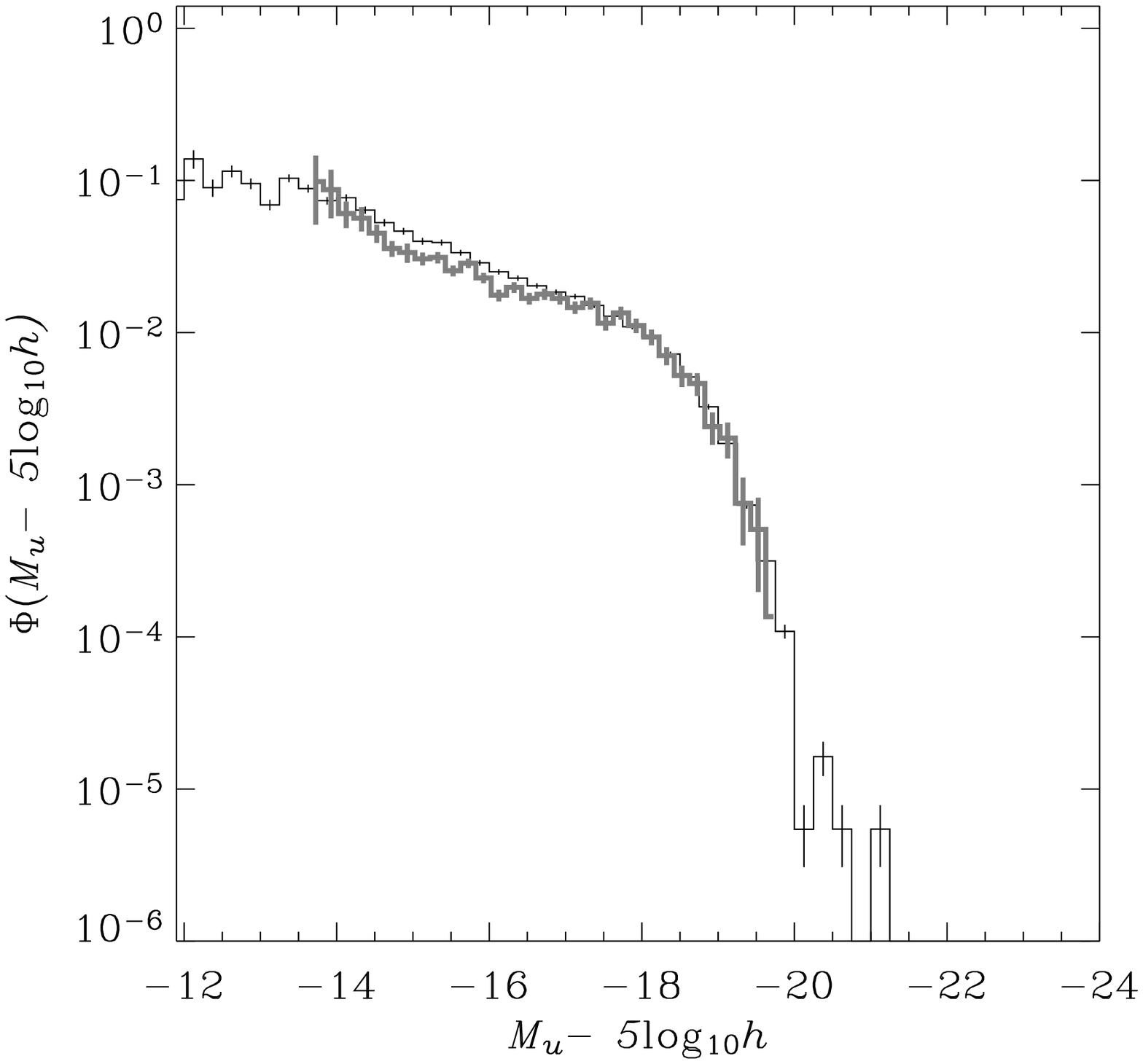}
\caption{\label{lf_u} Similar to Figure \ref{lf_cfact}, but showing
only the uncorrected results in the $u$-band 
(thin black) compared to the results of
\citet{baldry05a} (thick grey histogram). The results are similar,
though there is a 20\% discrepancy at the low luminosity end.}
\end{figure}

\citet{baldry05a} have measured the luminosity function
in the $u$ band using a deeper (but more heterogeneously selected)
SDSS sample and accounting for systematic errors in $u$-band sky
subtraction which we ignore here. Figure \ref{lf_u} compares our
result with their results in the redshift range $0.02 < z < 0.04$. The
results are similar, though at the low luminosity end there is about a
20\% discrepancy in the density found in our sample and in their
sample. The \citet{baldry05a} sample is restricted to the 275 square
degrees covered by the SDSS Southern Survey, which raises the
possibility that the difference is simply cosmic variance.

\begin{figure}
\figurenum{\fignum}
\plotone{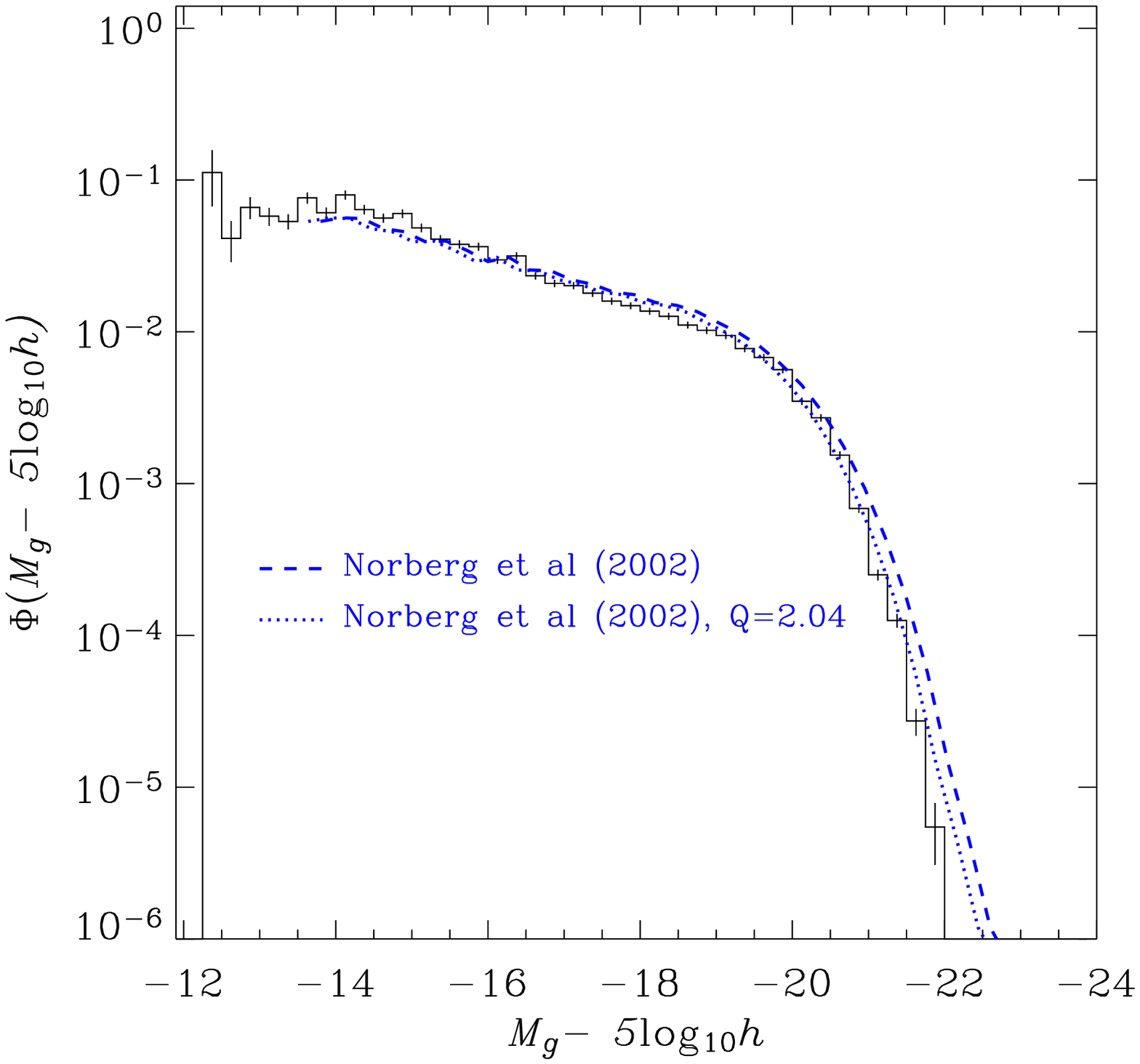}
\caption{\label{lf_g} Similar to Figure \ref{lf_cfact}, but showing
the uncorrected results for the step-wise maximum likelihood method in
the $g$-band (histogram) compared to the results of of
\citet{norberg02a} (dashed line). There is a fairly large discrepancy,
much of which can be explained by the weaker luminosity evolution
assumed by \citet{norberg02a}. The dotted line shows how the high
luminosity end would be shifted if we assume the stronger evolution
measured by \citet{blanton03c}. }
\end{figure}

In addition, Figure \ref{lf_g} compares the $g$ band luminosity
function to that of \citet{norberg02a}.  We convert the result of
\citet{norberg02a} in the $b_j$ band to $g$ using the relationship
$g=b_j-0.25$. The dashed line in Figure \ref{lf_g} shows the result of
converting their luminosity function this way. There are significant
differences between our $g$-band luminosity function and that of
\citet{norberg02a}. However, the slope at low luminosities is very
similar. Furthermore, we note that the luminosity evolution assumed by
\citet{norberg02a} is close to $Q=1.0$, much less than the empirical
luminosity evolution in the $g$-band $Q=2.04$ found by
\citet{blanton03c} and less than the theoretical estimate $Q=1.60$ of
\citet{bell03a}. An extremely simplistic way of accounting for this
difference at high luminosities is to simply shift their absolute
magnitudes by $\Delta Q z_m$, where $z_m=0.16$ is their median
redshift for galaxies with $M_g < -19.7$ and $\Delta Q = 1.04$ is the
difference between the effective evolution assumptions of
\citet{norberg02a} and the empirical estimates of
\citet{blanton04a}. This shift brings the results more closely in line
with each other at the luminous end, though fainter than that there
are still $20$--$30\%$ differences in the luminosity functions.

A real comparison may require a more careful understanding of the
different treatment of $K$-corrections and evolution in
\citet{norberg02a} and \citet{blanton03c}. $K$-corrections at
$z=0.25$, the limit of the sample of \citet{norberg02a}, are
significant; they can be up to 0.9 mag in the $g$ band. For the SDSS
bands, \citet{blanton03b} have shown that the $K$-corrections recover
fairly consistent intrinsic colors of galaxies as a function of
redshift, but this has not been shown with respect to the $b_j$ band,
and inconsistencies at the 10--20\% level have not been ruled out. The
effects of evolution are smaller, but much more poorly known;
theoretical estimates of the evolution do not agree with one another,
and the empirical measurement of \citet{blanton03c} is by no means
definitive.

The SDSS and the 2dFGRS are the most complete and well-studied
redshift surveys of the local universe, and aside from the roughly
0.15 mag disagreement in $M_\ast$ they are in fairly good agreement on
the overall shape of the luminosity function. There are no important
discrepancies in the low luminosity slope.

\subsection{Comparison between environments}

We can compare these field luminosity functions to those determined in
the SDSS from studies of clusters and voids. We leave for a future
paper the task of comparing the luminosity function in different
regions of our own sample, and here will concentrate on comparison to
the literature.

\begin{figure}
\figurenum{\fignum}
\plotone{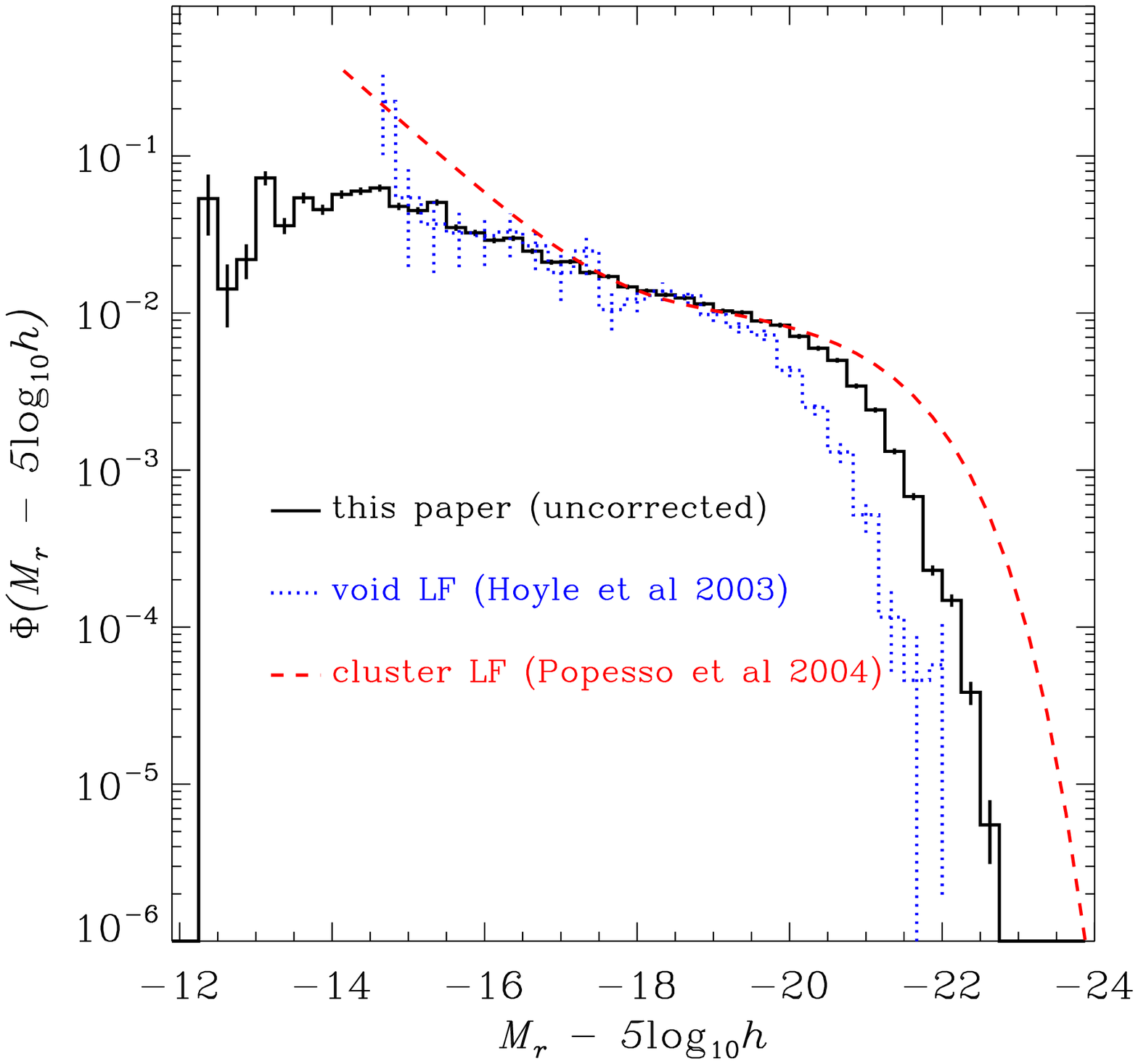}
\caption{\label{lf_cluster_void} Similar to Figure \ref{lf_cfact}, but showing
the uncorrected results for the step-wise maximum likelihood method in
the $r$-band (histogram) compared to the results for void galaxies
(\citealt{hoyle03a}) and for cluster galaxies
(\citealt{popesso04a}). }
\end{figure}

\citet{popesso04a} have used statistical background subtraction on the
photometric catalog to measure the optical luminosity function around
X-ray clusters without recourse to determining redshifts. Similar
methods have been used on other cluster samples by a number of
authors, such as \citet{schechter76a, gaidos97a, valotto97a,
garilli99a, paolillo01a, andreon02a, goto02a}. We will not attempt
here a comparison among all of those efforts. Figure
\ref{lf_cluster_void} shows the best-fit double Schechter function
result of \citet{popesso04a} compared to our raw luminosity function
(1). The exponential cut-off is clearly much brighter in clusters. In
addition, the low luminosity slope is much steeper below $M_r \sim
-17$. Note that the results of
\citet{popesso04a} should be free of the background subtraction biases
in cluster luminosity functions described by
\citet{valotto01a}, since the clusters are selected from X-ray
observations rather than the galaxy counts themselves. 

Both of the luminosity functions in Figure \ref{lf_cluster_void} are
susceptible to surface brightness selection effects. However, it is
{\it possible} that these effects are less important in the
\citet{popesso04a} results, since they do not apply the target
selection limit of $\mu_{50,r} < 24.5$ that affects the spectroscopic
results. 

Figure \ref{lf_cluster_void} also shows the luminosity function of
SDSS void galaxies found by \citet{hoyle03a}. Their luminosity
function is affected by exactly the same surface brightness limits as
this paper is. For these galaxies, there is a significantly less
luminous exponential cut-off, but the low luminosity slope is
remarkably similar to that of the field galaxy population.

\subsection{Luminosity function as a function of galaxy properties}
\label{props}

The luminosity function is known to be a function of galaxy type ---
among the many papers on this subject are those of
\citet{binggeli88a, roberts94a, marzke94a, marzke97a, bromley98a,
madgwick02a, blanton03d}. In this section, we explore this dependence
using the sample at hand.

\begin{figure}
\figurenum{\fignum}
\plotone{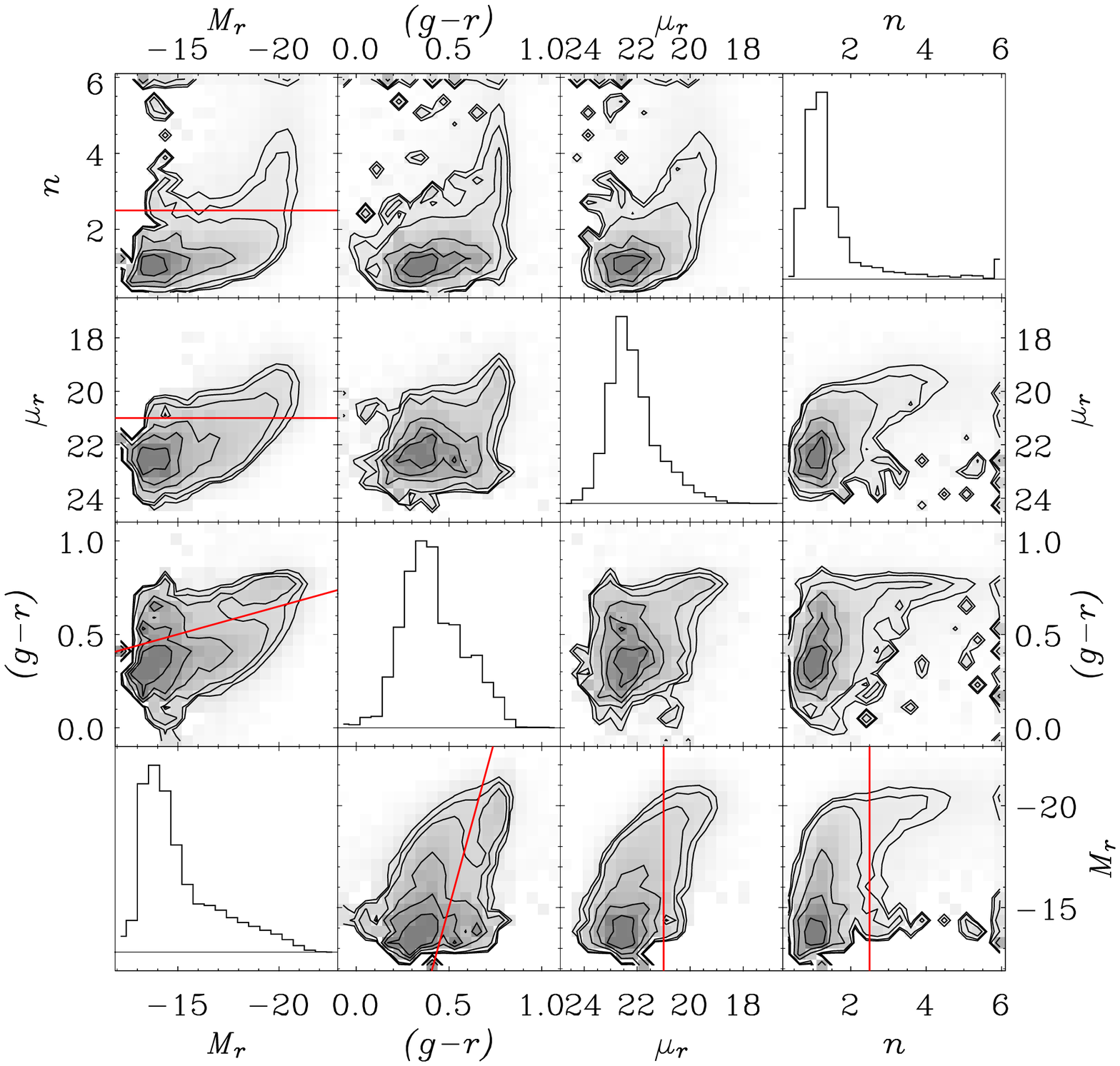}
\caption{\label{props_den} Number density distribution of galaxy
properties.  All images have a square-root stretch applied to increase
the dynamic range of the plot. Contours indicate the regions
containing 10\%, 30\%, 50\%, 70\%, 90\%, 95\%, and 97\% of the total
number of galaxies in this sample.  The upper and low triangles are
identical mirror images.  The histograms along the diagonal show the
density distribution of galaxies in each dimension.  We have weighted
each galaxy with $1/\Vmax$ to get the number density distribution of
galaxies. The lines indicate the cuts used in Figures
\ref{props_den_blue}--\ref{lf_cuts}. These densities have not been
corrected for surface brightness incompleteness in any way.}
\end{figure}

Figure \ref{props_den} shows the number density distribution of galaxy
properties (determined, in this case, using the 1/\Vmax\ estimator for
convenience). We have not corrected these plots for surface brightness
incompleteness in any way. The diagonal plots show the number density
distribution of absolute magnitude $M_r$, color $g-r$, $r$-band
half-light surface brightness $\mu_{50,r}$, and \Sersic\ index
$n$. The off-diagonal plots show the bivariate number distribution
between each pair of properties as contours and a greyscale. Many of
the qualitative features in this plot are extensions of the
relationships found in
\citet{blanton03d} for more luminous galaxies ($M_r < -17$). Low
luminosity galaxies are bluer, lower surface brightness, and more
exponential than high luminosity galaxies. Note that at low luminosity
(say $M_r > -16$ or so) the surface brightness selection at around
$\mu_{50,r} \sim 24$ clearly becomes important.

In the left column, one can see the luminosity function as a function
of color, surface brightness and \Sersic\ index.  The principal trend
which is evident is the large density of red, low luminosity galaxies.
We have seen above that these red, low luminosity galaxies are
preferentially in dense regions.

\begin{figure}
\figurenum{\fignum}
\plotone{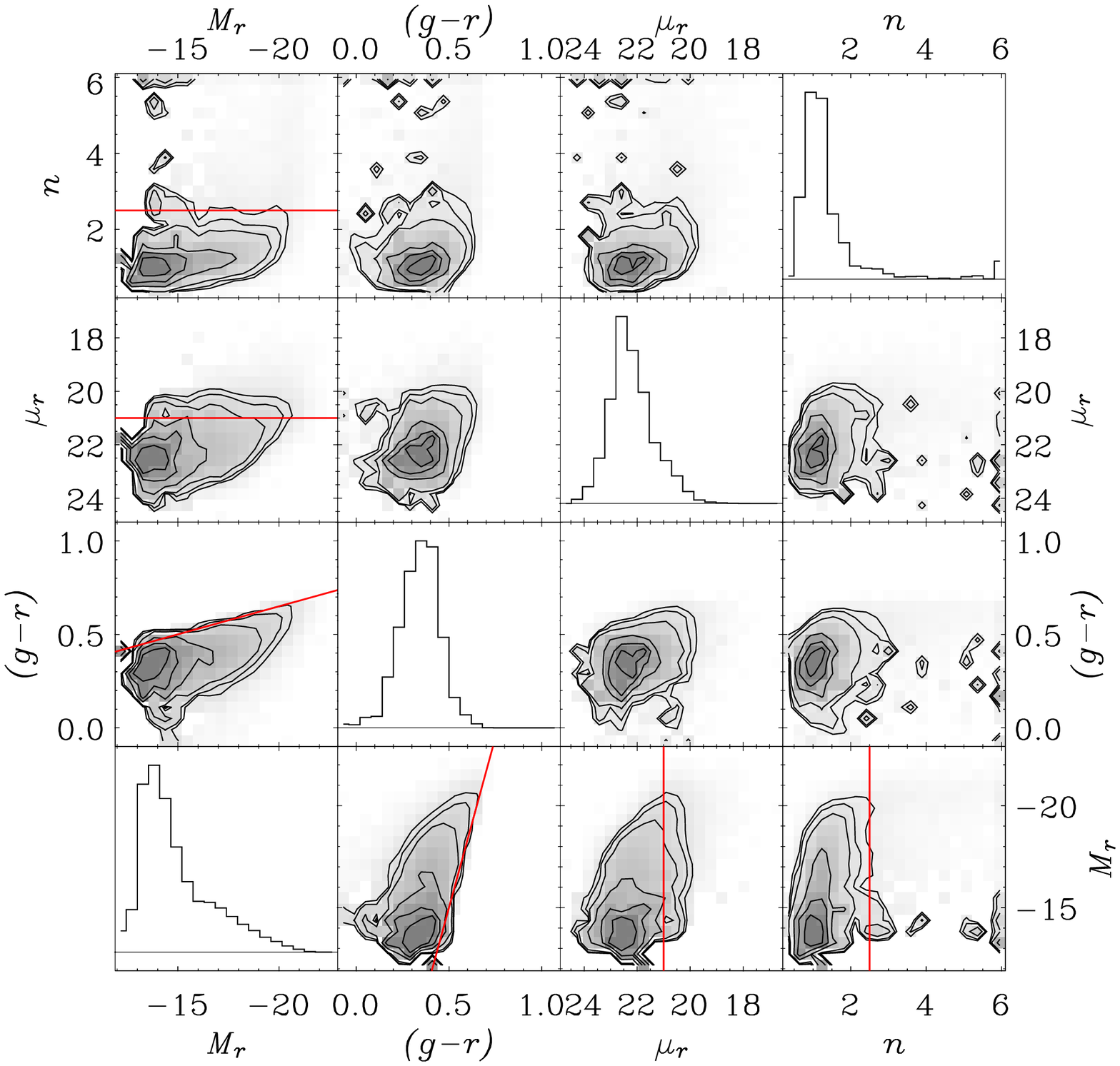}
\caption{\label{props_den_blue} Similar to Figure \ref{props_den}, but
for the blue galaxies (those bluer than the $(g-r)$ cut
shown). Generally these objects are low luminosity, exponential, and
low surface brightness.}
\end{figure}

\begin{figure}
\figurenum{\fignum}
\plotone{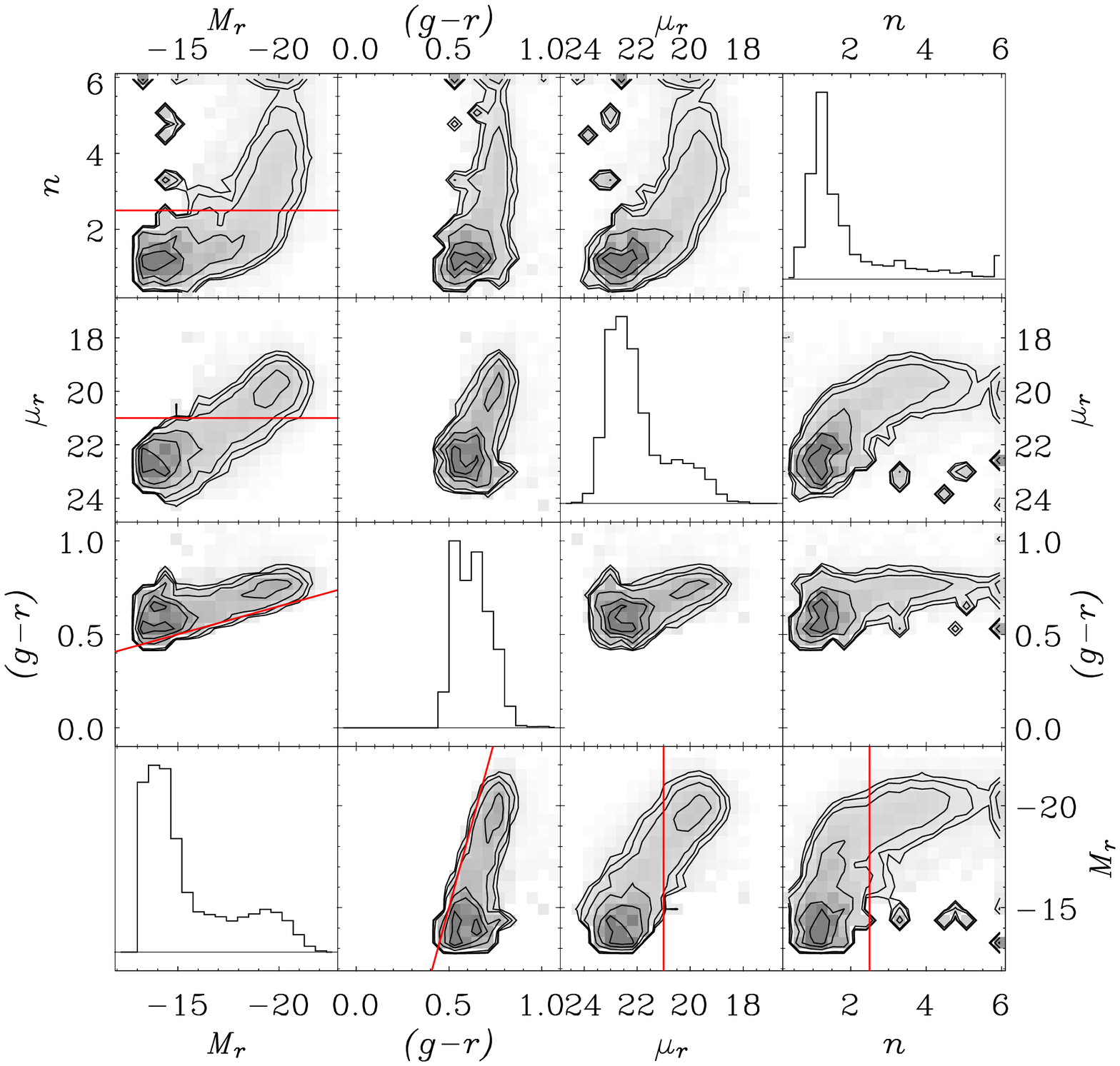}
\caption{\label{props_den_red} Similar to Figure \ref{props_den}, but
for the red galaxies (those redder than the $(g-r)$ cut shown). Note
the very strong relationships among absolute magnitude, surface
brightness, and \Sersic\ index for these red galaxies. }
\end{figure}

In order to investigate the properties of these red galaxies somewhat
further, we split the sample along color using the following,
luminosity-dependent cut:
\begin{equation}
\label{gmrcut}
(g-r)_c = 0.65 - 0.03 (M_r+20)
\end{equation}
This dividing line, shown as the tilted line in the appropriate panel
of Figure \ref{props_den}, is approximately at the bottom edge of the
red sequence.  Figure \ref{props_den_blue} shows the galaxies defined
as blue by this cut. Even at high luminosity, these tend to be low
surface brightness and not particularly concentrated.  Figure
\ref{props_den_red} shows the galaxies defined as red. The strong
dependence of \Sersic\ index and surface brightness on luminosity is
evident for the red galaxies. The lowest luminosity ones are as
unconcentrated and low surface brightness as the correspondingly
luminous blue galaxies, and generally are featureless and similar
morphologically to dE galaxies. In addition, one can easily see that
they have comparable space densities to more luminous red galaxies.

\begin{figure}
\figurenum{\fignum}
\epsscale{0.7}
\plotone{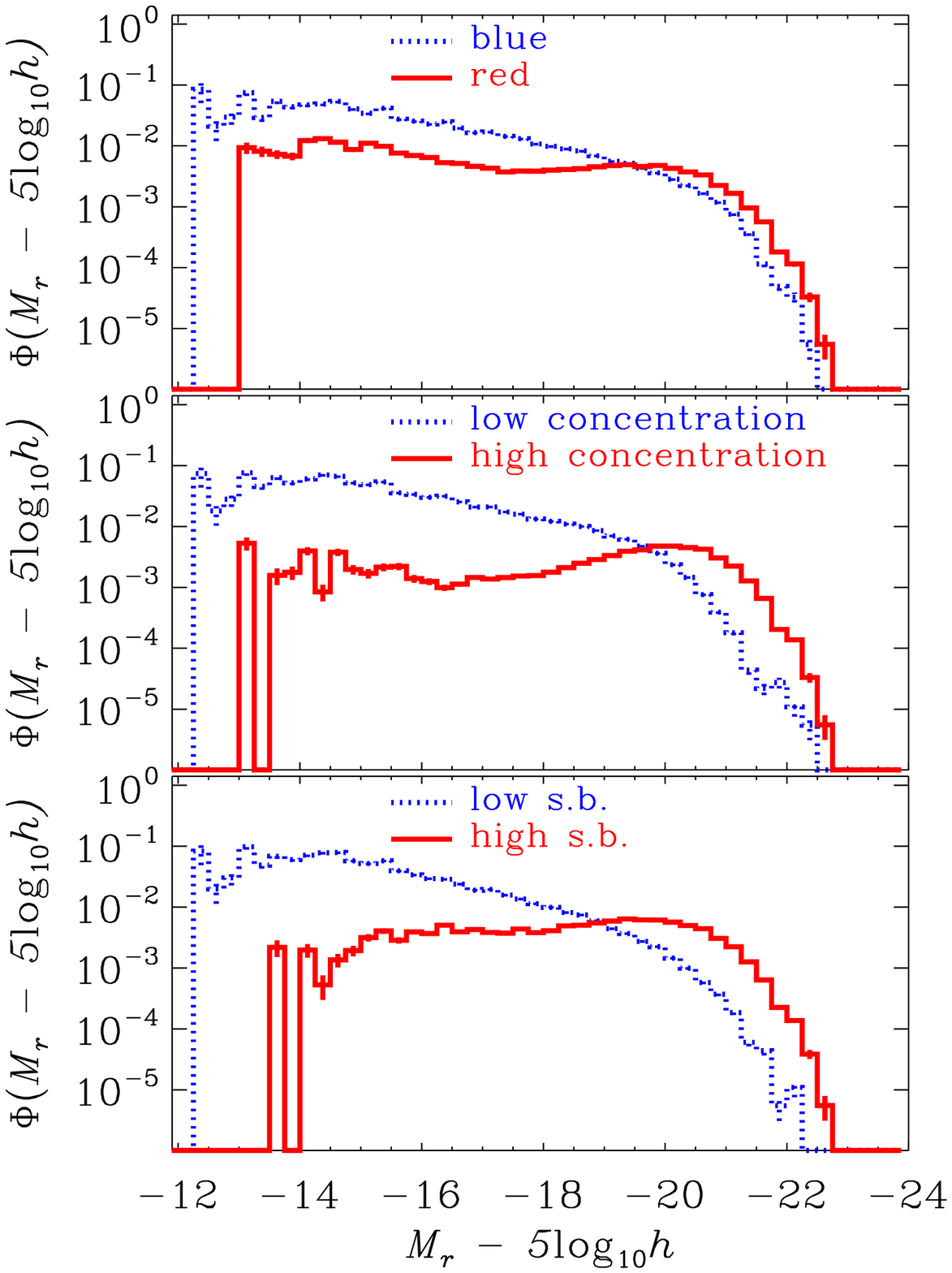}
\epsscale{1.0}
\vspace{0.4in} \caption{\label{lf_cuts} Raw luminosity function (1) found using the step-wise maximum likelihood method for galaxies split in three different ways:
using the $(g-r)$ cut of Equation \ref{gmrcut}, at $n=2.5$, and at
$\mu_{50,r}=21$ (the cuts shown as the lines in Figure
\ref{props_den}). }
\end{figure}

In order to better see the luminosity functions of galaxies of various
types, Figure \ref{lf_cuts} shows the raw luminosity function (1) for
galaxies split into two categories, in three different ways: by color,
as in Equation \ref{gmrcut}; by \Sersic\ index, dividing the
population at $n=2.5$; and by surface brightness, dividing the
population at $\mu_{50,r} = 21$.  We show the locations of these cuts
as lines in Figure \ref{props_den}. As in the previous plots, here we
see that a large fraction of low luminosity galaxies are blue, low
concentration, and low surface brightness.

\section{Summary}
\label{summary}

We have presented a sample and basic properties of a set of low
redshift ($z <0.05$) galaxies in the SDSS. Most interestingly, this
sample is the only sample of galaxies extending down to $M_r
\sim -12.5$ which is unbiased with respect to environment. 

From our measurements of the properties of these galaxies and of their
luminosity function, we conclude:
\begin{enumerate}
\item The slope of the luminosity function at low luminosities ($M_r -
5\log_{10} h > -17$), here denoted $\alpha_2$, is at least as steep as
$-1.3$, though it is almost certainly steeper, and perhaps as steep as
$-1.5$.
\item Galaxies at these low luminosities are low in surface
brightness, close to exponential, and predominantly blue.
\item Our estimate of the luminosity function is strongly
affected by surface brightness selection effects at low luminosities.
We have taken some care in estimating the completeness in this regime.
\item Incompleteness at low luminosities hardly affects our estimated
total luminosity density because even with our completeness
corrections the low lumosity galaxies do not contribute significantly.
\item There is not a large discrepancy between the SDSS and 2dFGRS
determinations of the luminosity function. 
\item Comparing void, field, and cluster luminosity functions in the
SDSS yields the conclusions that the exponential cut-off at high
luminosities is a strong function of environment and that $\alpha_2$
is not a function of environment at low density, but may become
significantly steeper at high density.
\end{enumerate}

There are a number of important improvements we can make to the
measurements described here:
\begin{enumerate}
\item With later data releases of the SDSS, we can explore more of the
local volume, of which we do not yet have a fair sample. In
particular, we will be observing regions surrounding and including the
Virgo cluster. This increased volume will allow a better measurement
of the environmental dependence of the properties of low luminosity
galaxies. 
\item Deeper SDSS spectroscopy to $m_r < 19.5$, targeted at galaxies
thought to be at $z<0.1$ according to photometric redshifts, may
result in improved statistics on the abundance of low luminosity
galaxies (Lin et al, in preparation). 
\item Improved analysis of the SDSS imaging will allow us to detect
low surface brightness galaxies. First, in the Southern Equatorial
stripe of the SDSS, we have many epochs of imaging, which will allow
us to detect galaxies by their diffuse light. Second,
\citet{willman01a} and \citet{zucker04a} continue to search for very
nearby galaxies detected in resolved stars. These samples will allow
us to evaluate the effects of surface brightness completeness on the
low luminosity slope empirically.
\item In order to improve the sample at the luminous end, we will
supplement the SDSS redshifts with known large, bright galaxies from
the astronomical literature, in a statistically meaningful way.
\end{enumerate}
These last two items will allow us to probe the regime below 10
$h^{-1}$ Mpc, which we have excluded here because of the current
incompleteness of the catalog in that regime. In addition to examining
even lower luminosity galaxies, we will be able to perform a more
direct comparison with the work of \citet{karachentsev04a} at these
distances.

With the data in hand, we will be studying further the environmental
dependence of galaxy properties, analyzing the images to better
understand the morphology of galaxies over a large range of
luminosities, and seeking to perform follow-up observations to better
understand the mass-to-light ratios, star-formation histories, and
dust content of extremely low luminosity galaxies in the field.
  
\acknowledgments

Thanks to Ivan Baldry, Eric Bell, Marla Geha, and Beth Willman for
comments on early drafts of the manuscript. Thanks to Julianne
Dalcanton, Rob Kennicutt, David Weinberg and Andrew West for useful
discussions and encouragement. Thanks to Mike Disney for suggesting
the test of Section \ref{dusttest}. This work would not have been
possible without the tremendous {\tt idlutils} library of software
developed by Doug Finkbeiner, Scott Burles, and DJS, and the Goddard
distribution of software distributed by Wayne Landsman. We only wish
there were publications to cite for these tools. MB acknowledges NASA
NAG5-11669 for partial support. MAS acknowledges the support of NSF
grant AST-0307409.

Funding for the creation and distribution of the SDSS Archive has been
provided by the Alfred P. Sloan Foundation, the Participating
Institutions, the National Aeronautics and Space Administration, the
National Science Foundation, the U.S. Department of Energy, the
Japanese Monbukagakusho, and the Max Planck Society. The SDSS Web site
is http://www.sdss.org/.

The SDSS is managed by the Astrophysical Research Consortium (ARC) for
the Participating Institutions. The Participating Institutions are The
University of Chicago, Fermilab, the Institute for Advanced Study, the
Japan Participation Group, The Johns Hopkins University, the Korean
Scientist Group, Los Alamos National Laboratory, the
Max-Planck-Institute for Astronomy (MPIA), the Max-Planck-Institute
for Astrophysics (MPA), New Mexico State University, University of
Pittsburgh, University of Portsmouth, Princeton University, the United
States Naval Observatory, and the University of Washington.

\newpage

\clearpage
\clearpage

\setcounter{thefigs}{0}

\clearpage
\stepcounter{thefigs}

\clearpage

\clearpage
\stepcounter{thefigs}

\clearpage
\stepcounter{thefigs}

\clearpage
\stepcounter{thefigs}

\clearpage
\stepcounter{thefigs}

\clearpage
\stepcounter{thefigs}

\clearpage
\stepcounter{thefigs}

\clearpage
\stepcounter{thefigs}

\clearpage
\stepcounter{thefigs}

\clearpage
\stepcounter{thefigs}

\clearpage
\stepcounter{thefigs}

\clearpage
\stepcounter{thefigs}

\clearpage
\stepcounter{thefigs}

\clearpage
\stepcounter{thefigs}

\clearpage
\stepcounter{thefigs}
\begin{figure}
\figurenum{\fignum}
\plotone{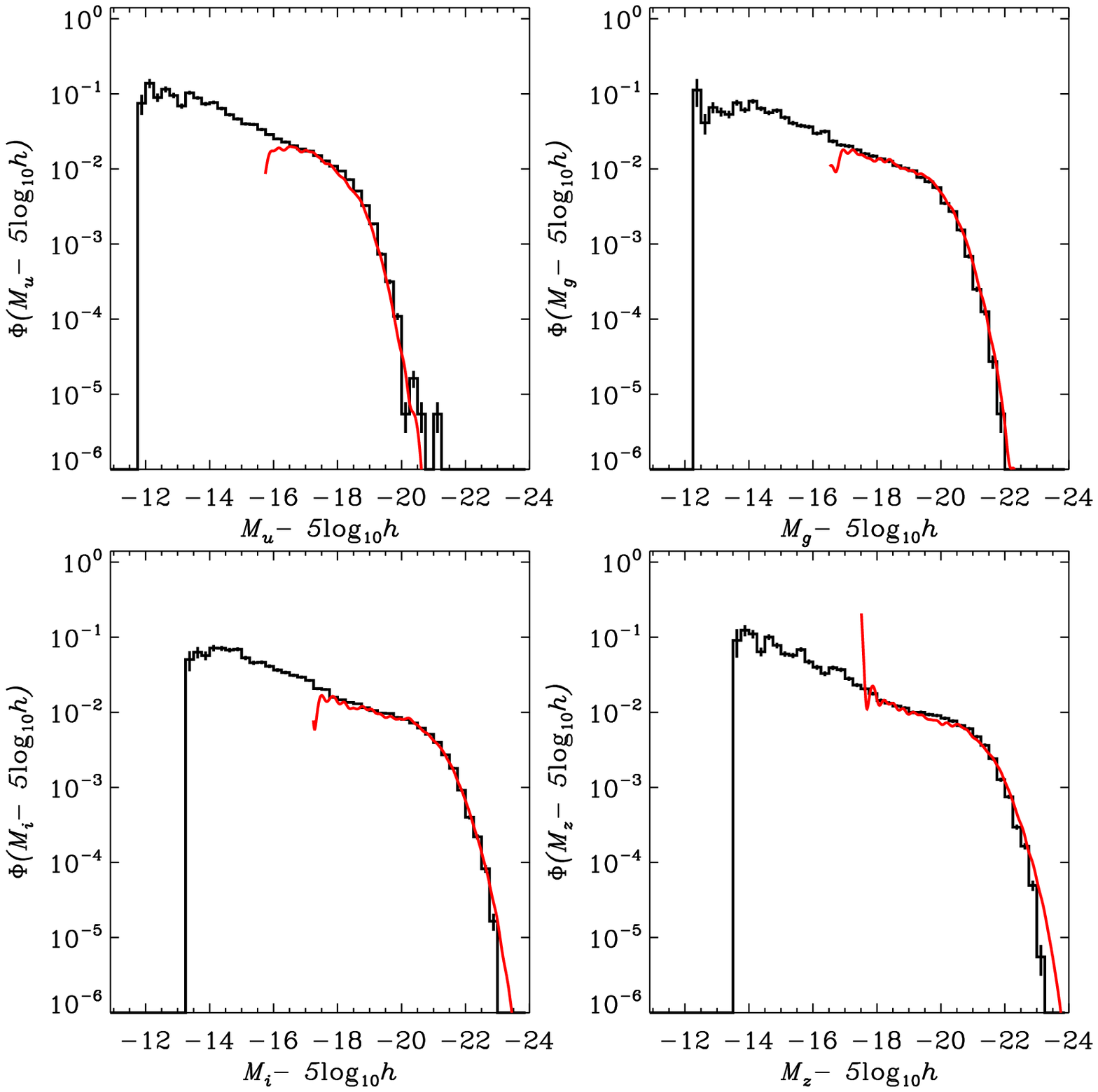}
\caption{\label{lf_bands_old} The histogram in each panel is
identical to those in Figure \ref{lf_bands}.  The smooth solid line is
the corresponding result from \citet{blanton03c}, evolution and
$K$-corrected to $z=0$. }
\end{figure}

\clearpage
\stepcounter{thefigs}

\clearpage
\stepcounter{thefigs}

\clearpage
\stepcounter{thefigs}

\clearpage
\stepcounter{thefigs}

\clearpage
\stepcounter{thefigs}

\clearpage
\stepcounter{thefigs}

\newpage
\clearpage
\begin{deluxetable}{rrrrrr}
\tablewidth{0pt}
\tablecolumns{6}
\tablecaption{\label{totalsb_table} Surface brightness completeness}
\tablehead{ $\mu_{50,r}$ & $N$ & $f_{\mathrm{ph}}(\mu_{50})$ &
  $\tilde{f}_{\mathrm{ti}}(\mu_{50})$ & $f_{\mathrm{sp}}(\mu_{50})$ & $f(\mu_{50})$ }
\startdata
18.16 & 32 & 0.78 & 0.82 & 1.00 & 0.64 \cr
18.49 & 144 & 0.86 & 0.93 & 1.00 & 0.80 \cr
18.81 & 344 & 0.95 & 0.98 & 1.00 & 0.93 \cr
19.14 & 809 & 0.98 & 1.01 & 1.00 & 0.99 \cr
19.46 & 1351 & 0.99 & 1.01 & 1.00 & 0.99 \cr
19.79 & 1747 & 0.97 & 1.01 & 1.00 & 0.98 \cr
20.11 & 2082 & 0.98 & 1.01 & 1.00 & 0.98 \cr
20.44 & 2480 & 0.98 & 1.01 & 1.00 & 0.98 \cr
20.76 & 2557 & 0.98 & 1.01 & 1.00 & 0.99 \cr
21.09 & 2672 & 0.97 & 1.00 & 1.00 & 0.97 \cr
21.41 & 2554 & 0.96 & 1.00 & 1.00 & 0.96 \cr
21.74 & 2334 & 0.96 & 0.99 & 1.00 & 0.94 \cr
22.06 & 1853 & 0.97 & 0.97 & 1.00 & 0.94 \cr
22.39 & 1469 & 0.94 & 0.95 & 1.00 & 0.88 \cr
22.71 & 940 & 0.86 & 0.89 & 0.99 & 0.76 \cr
23.04 & 490 & 0.84 & 0.87 & 0.96 & 0.71 \cr
23.36 & 165 & 0.76 & 0.72 & 0.91 & 0.50 \cr
23.69 & 47 & 0.63 & 0.68 & 0.78 & 0.33 \cr
24.01 & 8 & 0.44 & 0.38 & 0.56 & 0.09 \cr
24.34 & 5 & 0.33 & 0.00 & 0.00 & 0.00 \cr
\enddata
\tablecomments{As described in Section \ref{fgot}, the completeness as
  a function of surface brightness in any given direction $(\alpha,
  \delta)$ is the final column of this table multiplied by the window
  function $f_{\mathrm{got}}(\alpha, \delta)$ from
  \citet{blanton04a}. Note that $\tilde{f}_{\mathrm{ti}}(\mu_{50}) >1$
  in some cases, because it has had the overall tiling completeness
  divided out, as described in Section \ref{ti_fgot}.}
\end{deluxetable}

\begin{deluxetable}{rrrrr}
\tablewidth{0pt}
\tablecolumns{5}
\tablecaption{\label{lf_cfact_table} $r$-band luminosity function}
\tabletypesize{\small}
\tablehead{ $M_r - 5 \log_{10} h$ & $\Phi(M_r - 5 \log_{10} h)$ &
$\Phi(M_r - 5 \log_{10} h)$ & $\Phi(M_r - 5 \log_{10} h)$ 
& $\Phi(M_r - 5 \log_{10} h)$ \cr
& (raw, $\mu_{50,r} < 24$) & (corrected, $\mu_{50,r} < 24$) & (``total'') & 
(uncorrected, \Vmax\ method)}
\startdata
$-12.375$ & $(5.36 \pm 2.24) \times 10^{-2}$ & $(6.25 \pm 2.73) \times 10^{-2}$ & $(21.35 \pm 8.94) \times 10^{-2}$ & $(1.59 \pm 1.59) \times 10^{-2}$ \cr
$-12.625$ & $(1.42 \pm 0.61) \times 10^{-2}$ & $(1.65 \pm 0.73) \times 10^{-2}$ & $(5.12 \pm 2.21) \times 10^{-2}$ & $(9.49 \pm 0.95) \times 10^{-3}$ \cr
$-12.875$ & $(2.19 \pm 0.54) \times 10^{-2}$ & $(2.72 \pm 0.68) \times 10^{-2}$ & $(7.19 \pm 1.79) \times 10^{-2}$ & $(2.23 \pm 1.29) \times 10^{-2}$ \cr
$-13.125$ & $(7.25 \pm 0.73) \times 10^{-2}$ & $(10.52 \pm 1.03) \times 10^{-2}$ & $(21.87 \pm 2.22) \times 10^{-2}$ & $(9.57 \pm 2.27) \times 10^{-2}$ \cr
$-13.375$ & $(3.60 \pm 0.42) \times 10^{-2}$ & $(4.99 \pm 0.57) \times 10^{-2}$ & $(10.00 \pm 1.17) \times 10^{-2}$ & $(5.14 \pm 1.38) \times 10^{-2}$ \cr
$-13.625$ & $(5.41 \pm 0.42) \times 10^{-2}$ & $(7.59 \pm 0.58) \times 10^{-2}$ & $(13.89 \pm 1.09) \times 10^{-2}$ & $(8.97 \pm 1.57) \times 10^{-2}$ \cr
$-13.875$ & $(4.56 \pm 0.33) \times 10^{-2}$ & $(5.99 \pm 0.44) \times 10^{-2}$ & $(10.82 \pm 0.79) \times 10^{-2}$ & $(7.48 \pm 1.20) \times 10^{-2}$ \cr
$-14.125$ & $(5.69 \pm 0.34) \times 10^{-2}$ & $(8.07 \pm 0.48) \times 10^{-2}$ & $(12.55 \pm 0.76) \times 10^{-2}$ & $(8.52 \pm 1.09) \times 10^{-2}$ \cr
$-14.375$ & $(5.98 \pm 0.33) \times 10^{-2}$ & $(7.73 \pm 0.42) \times 10^{-2}$ & $(12.26 \pm 0.67) \times 10^{-2}$ & $(7.72 \pm 0.87) \times 10^{-2}$ \cr
$-14.625$ & $(6.25 \pm 0.32) \times 10^{-2}$ & $(8.26 \pm 0.41) \times 10^{-2}$ & $(11.97 \pm 0.61) \times 10^{-2}$ & $(6.87 \pm 0.69) \times 10^{-2}$ \cr
$-14.875$ & $(4.78 \pm 0.25) \times 10^{-2}$ & $(6.56 \pm 0.34) \times 10^{-2}$ & $(8.57 \pm 0.45) \times 10^{-2}$ & $(4.53 \pm 0.47) \times 10^{-2}$ \cr
$-15.125$ & $(4.49 \pm 0.22) \times 10^{-2}$ & $(6.07 \pm 0.30) \times 10^{-2}$ & $(7.57 \pm 0.38) \times 10^{-2}$ & $(3.88 \pm 0.37) \times 10^{-2}$ \cr
$-15.375$ & $(5.06 \pm 0.23) \times 10^{-2}$ & $(6.66 \pm 0.30) \times 10^{-2}$ & $(8.06 \pm 0.36) \times 10^{-2}$ & $(3.89 \pm 0.31) \times 10^{-2}$ \cr
$-15.625$ & $(3.50 \pm 0.17) \times 10^{-2}$ & $(4.55 \pm 0.21) \times 10^{-2}$ & $(5.28 \pm 0.25) \times 10^{-2}$ & $(2.45 \pm 0.21) \times 10^{-2}$ \cr
$-15.875$ & $(3.24 \pm 0.14) \times 10^{-2}$ & $(3.95 \pm 0.18) \times 10^{-2}$ & $(4.65 \pm 0.21) \times 10^{-2}$ & $(2.16 \pm 0.17) \times 10^{-2}$ \cr
$-16.125$ & $(2.91 \pm 0.12) \times 10^{-2}$ & $(3.60 \pm 0.15) \times 10^{-2}$ & $(3.98 \pm 0.17) \times 10^{-2}$ & $(2.14 \pm 0.14) \times 10^{-2}$ \cr
$-16.375$ & $(2.99 \pm 0.12) \times 10^{-2}$ & $(3.58 \pm 0.14) \times 10^{-2}$ & $(3.93 \pm 0.15) \times 10^{-2}$ & $(2.31 \pm 0.12) \times 10^{-2}$ \cr
$-16.625$ & $(2.48 \pm 0.09) \times 10^{-2}$ & $(2.88 \pm 0.11) \times 10^{-2}$ & $(3.13 \pm 0.12) \times 10^{-2}$ & $(2.02 \pm 0.10) \times 10^{-2}$ \cr
$-16.875$ & $(2.11 \pm 0.08) \times 10^{-2}$ & $(2.47 \pm 0.09) \times 10^{-2}$ & $(2.58 \pm 0.09) \times 10^{-2}$ & $(1.88 \pm 0.08) \times 10^{-2}$ \cr
$-17.125$ & $(2.12 \pm 0.07) \times 10^{-2}$ & $(2.46 \pm 0.09) \times 10^{-2}$ & $(2.52 \pm 0.09) \times 10^{-2}$ & $(2.05 \pm 0.07) \times 10^{-2}$ \cr
$-17.375$ & $(1.81 \pm 0.06) \times 10^{-2}$ & $(2.07 \pm 0.07) \times 10^{-2}$ & $(2.09 \pm 0.07) \times 10^{-2}$ & $(1.81 \pm 0.06) \times 10^{-2}$ \cr
$-17.625$ & $(1.71 \pm 0.06) \times 10^{-2}$ & $(1.91 \pm 0.06) \times 10^{-2}$ & $(1.93 \pm 0.07) \times 10^{-2}$ & $(1.64 \pm 0.05) \times 10^{-2}$ \cr
$-17.875$ & $(1.47 \pm 0.05) \times 10^{-2}$ & $(1.63 \pm 0.05) \times 10^{-2}$ & $(1.63 \pm 0.06) \times 10^{-2}$ & $(1.40 \pm 0.04) \times 10^{-2}$ \cr
$-18.125$ & $(1.38 \pm 0.05) \times 10^{-2}$ & $(1.51 \pm 0.05) \times 10^{-2}$ & $(1.51 \pm 0.05) \times 10^{-2}$ & $(1.37 \pm 0.03) \times 10^{-2}$ \cr
$-18.375$ & $(1.30 \pm 0.04) \times 10^{-2}$ & $(1.40 \pm 0.04) \times 10^{-2}$ & $(1.40 \pm 0.05) \times 10^{-2}$ & $(1.31 \pm 0.03) \times 10^{-2}$ \cr
$-18.625$ & $(1.25 \pm 0.04) \times 10^{-2}$ & $(1.32 \pm 0.04) \times 10^{-2}$ & $(1.33 \pm 0.04) \times 10^{-2}$ & $(1.25 \pm 0.03) \times 10^{-2}$ \cr
$-18.875$ & $(1.14 \pm 0.04) \times 10^{-2}$ & $(1.20 \pm 0.04) \times 10^{-2}$ & $(1.20 \pm 0.04) \times 10^{-2}$ & $(1.14 \pm 0.02) \times 10^{-2}$ \cr
$-19.125$ & $(1.03 \pm 0.04) \times 10^{-2}$ & $(1.07 \pm 0.03) \times 10^{-2}$ & $(1.08 \pm 0.04) \times 10^{-2}$ & $(1.03 \pm 0.02) \times 10^{-2}$ \cr
$-19.375$ & $(1.01 \pm 0.03) \times 10^{-2}$ & $(1.04 \pm 0.03) \times 10^{-2}$ & $(1.05 \pm 0.04) \times 10^{-2}$ & $(1.01 \pm 0.02) \times 10^{-2}$ \cr
$-19.625$ & $(8.89 \pm 0.30) \times 10^{-3}$ & $(9.18 \pm 0.29) \times 10^{-3}$ & $(9.19 \pm 0.31) \times 10^{-3}$ & $(8.91 \pm 0.22) \times 10^{-3}$ \cr
$-19.875$ & $(8.37 \pm 0.29) \times 10^{-3}$ & $(8.65 \pm 0.28) \times 10^{-3}$ & $(8.61 \pm 0.29) \times 10^{-3}$ & $(8.39 \pm 0.21) \times 10^{-3}$ \cr
$-20.125$ & $(7.10 \pm 0.25) \times 10^{-3}$ & $(7.27 \pm 0.23) \times 10^{-3}$ & $(7.28 \pm 0.25) \times 10^{-3}$ & $(7.12 \pm 0.20) \times 10^{-3}$ \cr
$-20.375$ & $(5.96 \pm 0.21) \times 10^{-3}$ & $(6.14 \pm 0.20) \times 10^{-3}$ & $(6.09 \pm 0.21) \times 10^{-3}$ & $(5.98 \pm 0.18) \times 10^{-3}$ \cr
$-20.625$ & $(5.00 \pm 0.18) \times 10^{-3}$ & $(5.13 \pm 0.17) \times 10^{-3}$ & $(5.09 \pm 0.18) \times 10^{-3}$ & $(5.01 \pm 0.17) \times 10^{-3}$ \cr
$-20.875$ & $(3.43 \pm 0.13) \times 10^{-3}$ & $(3.51 \pm 0.12) \times 10^{-3}$ & $(3.48 \pm 0.13) \times 10^{-3}$ & $(3.43 \pm 0.14) \times 10^{-3}$ \cr
$-21.125$ & $(2.42 \pm 0.09) \times 10^{-3}$ & $(2.48 \pm 0.09) \times 10^{-3}$ & $(2.44 \pm 0.09) \times 10^{-3}$ & $(2.42 \pm 0.11) \times 10^{-3}$ \cr
$-21.375$ & $(1.32 \pm 0.06) \times 10^{-3}$ & $(1.35 \pm 0.06) \times 10^{-3}$ & $(1.33 \pm 0.06) \times 10^{-3}$ & $(1.32 \pm 0.08) \times 10^{-3}$ \cr
$-21.625$ & $(6.78 \pm 0.34) \times 10^{-4}$ & $(6.89 \pm 0.34) \times 10^{-4}$ & $(6.81 \pm 0.35) \times 10^{-4}$ & $(6.78 \pm 0.61) \times 10^{-4}$ \cr
$-21.875$ & $(2.30 \pm 0.17) \times 10^{-4}$ & $(2.33 \pm 0.17) \times 10^{-4}$ & $(2.30 \pm 0.17) \times 10^{-4}$ & $(2.30 \pm 0.35) \times 10^{-4}$ \cr
$-22.125$ & $(1.48 \pm 0.13) \times 10^{-4}$ & $(1.52 \pm 0.14) \times 10^{-4}$ & $(1.48 \pm 0.13) \times 10^{-4}$ & $(1.48 \pm 0.28) \times 10^{-4}$ \cr
$-22.375$ & $(3.84 \pm 0.64) \times 10^{-5}$ & $(3.90 \pm 0.66) \times 10^{-5}$ & $(3.82 \pm 0.64) \times 10^{-5}$ & $(3.84 \pm 1.45) \times 10^{-5}$ \cr
$-22.625$ & $(5.50 \pm 2.40) \times 10^{-6}$ & $(5.54 \pm 2.44) \times 10^{-6}$ & $(5.45 \pm 2.37) \times 10^{-6}$ & $(5.51 \pm 5.51) \times 10^{-6}$ \cr
\enddata
\tablecomments{Column 2 is the raw luminosity function
  for galaxies with $\mu_{50,r} < 24$, corresponding to luminosity
  function (1) in Section \ref{sbcorr}. Column 3 is the luminosity
  function of galaxies with $\mu_{50,r} < 24$ corrected for surface
  brightness incompleteness, corresponding to luminosity function (2)
  in Section \ref{sbcorr}. Column 4 is the ``total'' luminosity
  function using the method described in Section \ref{sbcorr} to
  ``correct'' for the estimated missing fraction, corresponding to
  luminosity function (3) in that section. Column 5 is deprecated and
  is only given so the reader can evaluate the large-scale structure
  effects in the $1/V_{\mathrm{max}}$ weighted plots of Figures
  \ref{props_den}--\ref{props_den_red}. }
\end{deluxetable}

\begin{deluxetable}{rrrrr}
\tablewidth{0pt}
\tablecolumns{5}
\tablecaption{\label{lf_ugiz_table} $ugiz$-band luminosity functions}
\tabletypesize{\small}
\tablehead{ $M - 5\log_{10} h$ & $\Phi(M_u - 5\log_{10} h)$ & $\Phi(M_g - 5\log_{10} h)$ & $\Phi(M_i - 5\log_{10} h)$ &
  $\Phi(M_z - 5\log_{10} h)$}\startdata
$-11.875$ & 
$(7.49 \pm 2.23) \times 10^{-2}$ &
 --- &
 --- &
 --- \cr
$-12.125$ & 
$(1.39 \pm 0.19) \times 10^{-1}$ &
 --- &
 --- &
 --- \cr
$-12.375$ & 
$(8.97 \pm 1.15) \times 10^{-2}$ &
$(1.12 \pm 0.45) \times 10^{-1}$ &
 --- &
 --- \cr
$-12.625$ & 
$(1.15 \pm 0.10) \times 10^{-1}$ &
$(4.13 \pm 1.25) \times 10^{-2}$ &
 --- &
 --- \cr
$-12.875$ & 
$(9.55 \pm 0.76) \times 10^{-2}$ &
$(6.61 \pm 1.09) \times 10^{-2}$ &
 --- &
 --- \cr
$-13.125$ & 
$(6.91 \pm 0.54) \times 10^{-2}$ &
$(5.78 \pm 0.78) \times 10^{-2}$ &
 --- &
 --- \cr
$-13.375$ & 
$(1.04 \pm 0.06) \times 10^{-1}$ &
$(5.34 \pm 0.62) \times 10^{-2}$ &
$(5.03 \pm 1.50) \times 10^{-2}$ &
 --- \cr
$-13.625$ & 
$(8.85 \pm 0.51) \times 10^{-2}$ &
$(7.63 \pm 0.64) \times 10^{-2}$ &
$(6.31 \pm 1.08) \times 10^{-2}$ &
$(9.07 \pm 3.72) \times 10^{-2}$ \cr
$-13.875$ & 
$(7.37 \pm 0.42) \times 10^{-2}$ &
$(6.10 \pm 0.49) \times 10^{-2}$ &
$(5.69 \pm 0.76) \times 10^{-2}$ &
$(1.24 \pm 0.21) \times 10^{-1}$ \cr
$-14.125$ & 
$(7.70 \pm 0.41) \times 10^{-2}$ &
$(7.99 \pm 0.55) \times 10^{-2}$ &
$(7.17 \pm 0.68) \times 10^{-2}$ &
$(1.10 \pm 0.14) \times 10^{-1}$ \cr
$-14.375$ & 
$(6.39 \pm 0.34) \times 10^{-2}$ &
$(6.39 \pm 0.44) \times 10^{-2}$ &
$(7.12 \pm 0.55) \times 10^{-2}$ &
$(6.40 \pm 0.85) \times 10^{-2}$ \cr
$-14.625$ & 
$(5.29 \pm 0.28) \times 10^{-2}$ &
$(5.62 \pm 0.39) \times 10^{-2}$ &
$(6.74 \pm 0.45) \times 10^{-2}$ &
$(1.01 \pm 0.09) \times 10^{-1}$ \cr
$-14.875$ & 
$(4.65 \pm 0.24) \times 10^{-2}$ &
$(6.01 \pm 0.39) \times 10^{-2}$ &
$(6.92 \pm 0.41) \times 10^{-2}$ &
$(7.78 \pm 0.65) \times 10^{-2}$ \cr
$-15.125$ & 
$(3.99 \pm 0.20) \times 10^{-2}$ &
$(4.84 \pm 0.32) \times 10^{-2}$ &
$(5.29 \pm 0.32) \times 10^{-2}$ &
$(5.98 \pm 0.49) \times 10^{-2}$ \cr
$-15.375$ & 
$(3.91 \pm 0.18) \times 10^{-2}$ &
$(4.08 \pm 0.26) \times 10^{-2}$ &
$(4.57 \pm 0.28) \times 10^{-2}$ &
$(5.72 \pm 0.44) \times 10^{-2}$ \cr
$-15.625$ & 
$(3.35 \pm 0.15) \times 10^{-2}$ &
$(3.77 \pm 0.24) \times 10^{-2}$ &
$(4.64 \pm 0.26) \times 10^{-2}$ &
$(6.85 \pm 0.47) \times 10^{-2}$ \cr
$-15.875$ & 
$(2.88 \pm 0.12) \times 10^{-2}$ &
$(3.64 \pm 0.22) \times 10^{-2}$ &
$(4.10 \pm 0.22) \times 10^{-2}$ &
$(4.70 \pm 0.34) \times 10^{-2}$ \cr
$-16.125$ & 
$(2.51 \pm 0.10) \times 10^{-2}$ &
$(2.98 \pm 0.18) \times 10^{-2}$ &
$(3.66 \pm 0.19) \times 10^{-2}$ &
$(3.98 \pm 0.29) \times 10^{-2}$ \cr
$-16.375$ & 
$(2.28 \pm 0.09) \times 10^{-2}$ &
$(3.16 \pm 0.18) \times 10^{-2}$ &
$(3.38 \pm 0.17) \times 10^{-2}$ &
$(3.28 \pm 0.24) \times 10^{-2}$ \cr
$-16.625$ & 
$(2.03 \pm 0.08) \times 10^{-2}$ &
$(2.34 \pm 0.13) \times 10^{-2}$ &
$(3.11 \pm 0.15) \times 10^{-2}$ &
$(3.92 \pm 0.25) \times 10^{-2}$ \cr
$-16.875$ & 
$(1.85 \pm 0.07) \times 10^{-2}$ &
$(2.09 \pm 0.11) \times 10^{-2}$ &
$(2.94 \pm 0.13) \times 10^{-2}$ &
$(3.69 \pm 0.23) \times 10^{-2}$ \cr
$-17.125$ & 
$(1.73 \pm 0.07) \times 10^{-2}$ &
$(2.02 \pm 0.11) \times 10^{-2}$ &
$(2.66 \pm 0.11) \times 10^{-2}$ &
$(2.81 \pm 0.17) \times 10^{-2}$ \cr
$-17.375$ & 
$(1.51 \pm 0.06) \times 10^{-2}$ &
$(1.80 \pm 0.10) \times 10^{-2}$ &
$(2.07 \pm 0.08) \times 10^{-2}$ &
$(2.30 \pm 0.14) \times 10^{-2}$ \cr
$-17.625$ & 
$(1.28 \pm 0.05) \times 10^{-2}$ &
$(1.59 \pm 0.08) \times 10^{-2}$ &
$(2.01 \pm 0.08) \times 10^{-2}$ &
$(2.05 \pm 0.12) \times 10^{-2}$ \cr
$-17.875$ & 
$(1.09 \pm 0.04) \times 10^{-2}$ &
$(1.49 \pm 0.08) \times 10^{-2}$ &
$(1.59 \pm 0.06) \times 10^{-2}$ &
$(1.78 \pm 0.10) \times 10^{-2}$ \cr
$-18.125$ & 
$(9.37 \pm 0.36) \times 10^{-3}$ &
$(1.37 \pm 0.07) \times 10^{-2}$ &
$(1.46 \pm 0.05) \times 10^{-2}$ &
$(1.44 \pm 0.08) \times 10^{-2}$ \cr
$-18.375$ & 
$(7.23 \pm 0.28) \times 10^{-3}$ &
$(1.27 \pm 0.07) \times 10^{-2}$ &
$(1.35 \pm 0.05) \times 10^{-2}$ &
$(1.32 \pm 0.07) \times 10^{-2}$ \cr
$-18.625$ & 
$(5.11 \pm 0.20) \times 10^{-3}$ &
$(1.11 \pm 0.06) \times 10^{-2}$ &
$(1.30 \pm 0.05) \times 10^{-2}$ &
$(1.21 \pm 0.07) \times 10^{-2}$ \cr
$-18.875$ & 
$(3.26 \pm 0.13) \times 10^{-3}$ &
$(1.02 \pm 0.05) \times 10^{-2}$ &
$(1.15 \pm 0.04) \times 10^{-2}$ &
$(1.14 \pm 0.06) \times 10^{-2}$ \cr
$-19.125$ & 
$(1.87 \pm 0.08) \times 10^{-3}$ &
$(9.47 \pm 0.50) \times 10^{-3}$ &
$(1.05 \pm 0.04) \times 10^{-2}$ &
$(9.97 \pm 0.54) \times 10^{-3}$ \cr
$-19.375$ & 
$(7.35 \pm 0.38) \times 10^{-4}$ &
$(7.76 \pm 0.41) \times 10^{-3}$ &
$(9.77 \pm 0.33) \times 10^{-3}$ &
$(10.00 \pm 0.54) \times 10^{-3}$ \cr
$-19.625$ & 
$(3.16 \pm 0.21) \times 10^{-4}$ &
$(6.78 \pm 0.36) \times 10^{-3}$ &
$(9.63 \pm 0.33) \times 10^{-3}$ &
$(9.35 \pm 0.50) \times 10^{-3}$ \cr
$-19.875$ & 
$(1.09 \pm 0.11) \times 10^{-4}$ &
$(5.64 \pm 0.30) \times 10^{-3}$ &
$(8.57 \pm 0.29) \times 10^{-3}$ &
$(9.05 \pm 0.49) \times 10^{-3}$ \cr
$-20.125$ & 
$(5.44 \pm 2.37) \times 10^{-6}$ &
$(3.49 \pm 0.19) \times 10^{-3}$ &
$(8.02 \pm 0.28) \times 10^{-3}$ &
$(8.31 \pm 0.45) \times 10^{-3}$ \cr
$-20.375$ & 
$(1.63 \pm 0.41) \times 10^{-5}$ &
$(2.72 \pm 0.15) \times 10^{-3}$ &
$(7.21 \pm 0.25) \times 10^{-3}$ &
$(7.64 \pm 0.41) \times 10^{-3}$ \cr
$-20.625$ & 
$(5.44 \pm 2.37) \times 10^{-6}$ &
$(1.54 \pm 0.09) \times 10^{-3}$ &
$(6.15 \pm 0.22) \times 10^{-3}$ &
$(6.64 \pm 0.36) \times 10^{-3}$ \cr
$-20.875$ & 
 --- &
$(6.86 \pm 0.44) \times 10^{-4}$ &
$(5.10 \pm 0.18) \times 10^{-3}$ &
$(6.04 \pm 0.33) \times 10^{-3}$ \cr
$-21.125$ & 
$(5.45 \pm 2.37) \times 10^{-6}$ &
$(2.51 \pm 0.21) \times 10^{-4}$ &
$(3.97 \pm 0.14) \times 10^{-3}$ &
$(4.73 \pm 0.26) \times 10^{-3}$ \cr
$-21.375$ & 
 --- &
$(1.25 \pm 0.13) \times 10^{-4}$ &
$(2.72 \pm 0.10) \times 10^{-3}$ &
$(3.64 \pm 0.20) \times 10^{-3}$ \cr
$-21.625$ & 
 --- &
$(2.73 \pm 0.55) \times 10^{-5}$ &
$(1.79 \pm 0.07) \times 10^{-3}$ &
$(2.41 \pm 0.14) \times 10^{-3}$ \cr
$-21.875$ & 
 --- &
$(5.47 \pm 2.39) \times 10^{-6}$ &
$(9.19 \pm 0.43) \times 10^{-4}$ &
$(1.27 \pm 0.08) \times 10^{-3}$ \cr
$-22.125$ & 
 --- &
 --- &
$(4.00 \pm 0.24) \times 10^{-4}$ &
$(7.50 \pm 0.48) \times 10^{-4}$ \cr
$-22.375$ & 
 --- &
 --- &
$(2.19 \pm 0.17) \times 10^{-4}$ &
$(2.96 \pm 0.23) \times 10^{-4}$ \cr
$-22.625$ & 
 --- &
 --- &
$(8.24 \pm 0.96) \times 10^{-5}$ &
$(1.65 \pm 0.16) \times 10^{-4}$ \cr
$-22.875$ & 
 --- &
 --- &
$(1.65 \pm 0.42) \times 10^{-5}$ &
$(4.95 \pm 0.76) \times 10^{-5}$ \cr
$-23.125$ & 
 --- &
 --- &
 --- &
$(5.52 \pm 2.41) \times 10^{-6}$ \cr
\enddata
\vspace{-0.3in} \tablecomments{We have not corrected these luminosity functions
for surface brightness incompleteness at all; they correspond to
luminosity function (1) in Section \ref{sbcorr}.}
\end{deluxetable}

\begin{deluxetable}{rrrrr}
\tablewidth{0pt}
\tablecolumns{5}
\tablecaption{\label{lf_ugiz_2_table} $ugiz$-band luminosity
functions, surface brightness completeness corrected}
\tabletypesize{\small}
\tablehead{ $M - 5\log_{10} h$ & $\Phi(M_u - 5\log_{10} h)$ & $\Phi(M_g - 5\log_{10} h)$ & $\Phi(M_i - 5\log_{10} h)$ &
  $\Phi(M_z - 5\log_{10} h)$}
\startdata
$-11.875$ & 
$(1.19 \pm 0.34) \times 10^{-1}$ &
 --- &
 --- &
 --- \cr
$-12.125$ & 
$(1.93 \pm 0.26) \times 10^{-1}$ &
 --- &
 --- &
 --- \cr
$-12.375$ & 
$(1.30 \pm 0.16) \times 10^{-1}$ &
$(1.25 \pm 0.54) \times 10^{-1}$ &
 --- &
 --- \cr
$-12.625$ & 
$(1.61 \pm 0.14) \times 10^{-1}$ &
$(4.95 \pm 1.52) \times 10^{-2}$ &
 --- &
 --- \cr
$-12.875$ & 
$(1.20 \pm 0.10) \times 10^{-1}$ &
$(9.14 \pm 1.47) \times 10^{-2}$ &
 --- &
 --- \cr
$-13.125$ & 
$(9.83 \pm 0.77) \times 10^{-2}$ &
$(7.86 \pm 1.04) \times 10^{-2}$ &
 --- &
 --- \cr
$-13.375$ & 
$(1.40 \pm 0.08) \times 10^{-1}$ &
$(6.80 \pm 0.79) \times 10^{-2}$ &
$(5.68 \pm 1.74) \times 10^{-2}$ &
 --- \cr
$-13.625$ & 
$(1.21 \pm 0.07) \times 10^{-1}$ &
$(1.06 \pm 0.09) \times 10^{-1}$ &
$(1.08 \pm 0.17) \times 10^{-1}$ &
$(1.29 \pm 0.51) \times 10^{-1}$ \cr
$-13.875$ & 
$(9.93 \pm 0.59) \times 10^{-2}$ &
$(8.80 \pm 0.70) \times 10^{-2}$ &
$(6.66 \pm 0.94) \times 10^{-2}$ &
$(1.67 \pm 0.28) \times 10^{-1}$ \cr
$-14.125$ & 
$(9.59 \pm 0.52) \times 10^{-2}$ &
$(1.02 \pm 0.07) \times 10^{-1}$ &
$(8.14 \pm 0.81) \times 10^{-2}$ &
$(1.64 \pm 0.20) \times 10^{-1}$ \cr
$-14.375$ & 
$(8.33 \pm 0.46) \times 10^{-2}$ &
$(9.40 \pm 0.65) \times 10^{-2}$ &
$(1.02 \pm 0.08) \times 10^{-1}$ &
$(7.72 \pm 1.04) \times 10^{-2}$ \cr
$-14.625$ & 
$(6.04 \pm 0.33) \times 10^{-2}$ &
$(7.98 \pm 0.55) \times 10^{-2}$ &
$(9.80 \pm 0.64) \times 10^{-2}$ &
$(1.49 \pm 0.13) \times 10^{-1}$ \cr
$-14.875$ & 
$(5.91 \pm 0.31) \times 10^{-2}$ &
$(7.86 \pm 0.50) \times 10^{-2}$ &
$(9.74 \pm 0.57) \times 10^{-2}$ &
$(9.96 \pm 0.85) \times 10^{-2}$ \cr
$-15.125$ & 
$(4.60 \pm 0.24) \times 10^{-2}$ &
$(5.96 \pm 0.38) \times 10^{-2}$ &
$(6.89 \pm 0.42) \times 10^{-2}$ &
$(6.94 \pm 0.59) \times 10^{-2}$ \cr
$-15.375$ & 
$(4.44 \pm 0.21) \times 10^{-2}$ &
$(5.17 \pm 0.33) \times 10^{-2}$ &
$(5.35 \pm 0.33) \times 10^{-2}$ &
$(6.83 \pm 0.55) \times 10^{-2}$ \cr
$-15.625$ & 
$(3.84 \pm 0.17) \times 10^{-2}$ &
$(4.45 \pm 0.27) \times 10^{-2}$ &
$(5.91 \pm 0.33) \times 10^{-2}$ &
$(8.08 \pm 0.57) \times 10^{-2}$ \cr
$-15.875$ & 
$(3.22 \pm 0.14) \times 10^{-2}$ &
$(4.31 \pm 0.25) \times 10^{-2}$ &
$(4.84 \pm 0.27) \times 10^{-2}$ &
$(5.49 \pm 0.41) \times 10^{-2}$ \cr
$-16.125$ & 
$(2.76 \pm 0.12) \times 10^{-2}$ &
$(3.43 \pm 0.19) \times 10^{-2}$ &
$(4.22 \pm 0.22) \times 10^{-2}$ &
$(4.69 \pm 0.35) \times 10^{-2}$ \cr
$-16.375$ & 
$(2.51 \pm 0.10) \times 10^{-2}$ &
$(3.64 \pm 0.20) \times 10^{-2}$ &
$(3.85 \pm 0.19) \times 10^{-2}$ &
$(3.59 \pm 0.26) \times 10^{-2}$ \cr
$-16.625$ & 
$(2.20 \pm 0.09) \times 10^{-2}$ &
$(2.68 \pm 0.14) \times 10^{-2}$ &
$(3.51 \pm 0.17) \times 10^{-2}$ &
$(4.41 \pm 0.30) \times 10^{-2}$ \cr
$-16.875$ & 
$(1.98 \pm 0.08) \times 10^{-2}$ &
$(2.34 \pm 0.12) \times 10^{-2}$ &
$(3.33 \pm 0.15) \times 10^{-2}$ &
$(4.15 \pm 0.27) \times 10^{-2}$ \cr
$-17.125$ & 
$(1.83 \pm 0.07) \times 10^{-2}$ &
$(2.23 \pm 0.11) \times 10^{-2}$ &
$(2.95 \pm 0.12) \times 10^{-2}$ &
$(3.07 \pm 0.20) \times 10^{-2}$ \cr
$-17.375$ & 
$(1.59 \pm 0.06) \times 10^{-2}$ &
$(1.98 \pm 0.10) \times 10^{-2}$ &
$(2.27 \pm 0.09) \times 10^{-2}$ &
$(2.49 \pm 0.15) \times 10^{-2}$ \cr
$-17.625$ & 
$(1.34 \pm 0.05) \times 10^{-2}$ &
$(1.73 \pm 0.09) \times 10^{-2}$ &
$(2.20 \pm 0.08) \times 10^{-2}$ &
$(2.24 \pm 0.14) \times 10^{-2}$ \cr
$-17.875$ & 
$(1.13 \pm 0.04) \times 10^{-2}$ &
$(1.59 \pm 0.08) \times 10^{-2}$ &
$(1.74 \pm 0.07) \times 10^{-2}$ &
$(1.90 \pm 0.11) \times 10^{-2}$ \cr
$-18.125$ & 
$(9.68 \pm 0.37) \times 10^{-3}$ &
$(1.45 \pm 0.07) \times 10^{-2}$ &
$(1.57 \pm 0.06) \times 10^{-2}$ &
$(1.54 \pm 0.09) \times 10^{-2}$ \cr
$-18.375$ & 
$(7.44 \pm 0.29) \times 10^{-3}$ &
$(1.33 \pm 0.06) \times 10^{-2}$ &
$(1.45 \pm 0.05) \times 10^{-2}$ &
$(1.41 \pm 0.08) \times 10^{-2}$ \cr
$-18.625$ & 
$(5.25 \pm 0.21) \times 10^{-3}$ &
$(1.16 \pm 0.05) \times 10^{-2}$ &
$(1.38 \pm 0.05) \times 10^{-2}$ &
$(1.27 \pm 0.07) \times 10^{-2}$ \cr
$-18.875$ & 
$(3.34 \pm 0.14) \times 10^{-3}$ &
$(1.06 \pm 0.05) \times 10^{-2}$ &
$(1.21 \pm 0.04) \times 10^{-2}$ &
$(1.19 \pm 0.07) \times 10^{-2}$ \cr
$-19.125$ & 
$(1.91 \pm 0.08) \times 10^{-3}$ &
$(9.79 \pm 0.47) \times 10^{-3}$ &
$(1.10 \pm 0.04) \times 10^{-2}$ &
$(1.04 \pm 0.06) \times 10^{-2}$ \cr
$-19.375$ & 
$(7.52 \pm 0.40) \times 10^{-4}$ &
$(8.04 \pm 0.38) \times 10^{-3}$ &
$(1.01 \pm 0.03) \times 10^{-2}$ &
$(1.04 \pm 0.06) \times 10^{-2}$ \cr
$-19.625$ & 
$(3.26 \pm 0.22) \times 10^{-4}$ &
$(6.90 \pm 0.33) \times 10^{-3}$ &
$(9.96 \pm 0.34) \times 10^{-3}$ &
$(9.66 \pm 0.53) \times 10^{-3}$ \cr
$-19.875$ & 
$(3.84 \pm 0.45) \times 10^{-5}$ &
$(5.79 \pm 0.28) \times 10^{-3}$ &
$(8.83 \pm 0.30) \times 10^{-3}$ &
$(9.33 \pm 0.51) \times 10^{-3}$ \cr
$-20.125$ & 
$(5.52 \pm 2.42) \times 10^{-6}$ &
$(3.58 \pm 0.18) \times 10^{-3}$ &
$(8.20 \pm 0.28) \times 10^{-3}$ &
$(8.54 \pm 0.47) \times 10^{-3}$ \cr
$-20.375$ & 
$(1.66 \pm 0.42) \times 10^{-5}$ &
$(2.79 \pm 0.14) \times 10^{-3}$ &
$(7.44 \pm 0.26) \times 10^{-3}$ &
$(7.81 \pm 0.43) \times 10^{-3}$ \cr
$-20.625$ & 
$(5.53 \pm 2.42) \times 10^{-6}$ &
$(1.58 \pm 0.08) \times 10^{-3}$ &
$(6.36 \pm 0.22) \times 10^{-3}$ &
$(6.85 \pm 0.38) \times 10^{-3}$ \cr
$-20.875$ & 
 --- &
$(7.04 \pm 0.42) \times 10^{-4}$ &
$(5.23 \pm 0.18) \times 10^{-3}$ &
$(6.22 \pm 0.34) \times 10^{-3}$ \cr
$-21.125$ & 
$(7.03 \pm 2.92) \times 10^{-6}$ &
$(2.54 \pm 0.20) \times 10^{-4}$ &
$(4.06 \pm 0.15) \times 10^{-3}$ &
$(4.87 \pm 0.27) \times 10^{-3}$ \cr
$-21.375$ & 
 --- &
$(1.27 \pm 0.13) \times 10^{-4}$ &
$(2.80 \pm 0.11) \times 10^{-3}$ &
$(3.72 \pm 0.21) \times 10^{-3}$ \cr
$-21.625$ & 
 --- &
$(2.95 \pm 0.59) \times 10^{-5}$ &
$(1.84 \pm 0.07) \times 10^{-3}$ &
$(2.49 \pm 0.14) \times 10^{-3}$ \cr
$-21.875$ & 
 --- &
$(5.51 \pm 2.43) \times 10^{-6}$ &
$(9.43 \pm 0.44) \times 10^{-4}$ &
$(1.31 \pm 0.08) \times 10^{-3}$ \cr
$-22.125$ & 
 --- &
 --- &
$(4.06 \pm 0.24) \times 10^{-4}$ &
$(7.63 \pm 0.50) \times 10^{-4}$ \cr
$-22.375$ & 
 --- &
 --- &
$(2.24 \pm 0.17) \times 10^{-4}$ &
$(3.00 \pm 0.24) \times 10^{-4}$ \cr
$-22.625$ & 
 --- &
 --- &
$(8.39 \pm 0.98) \times 10^{-5}$ &
$(1.69 \pm 0.16) \times 10^{-4}$ \cr
$-22.875$ & 
 --- &
 --- &
$(1.67 \pm 0.42) \times 10^{-5}$ &
$(5.02 \pm 0.78) \times 10^{-5}$ \cr
$-23.125$ & 
 --- &
 --- &
 --- &
$(5.56 \pm 2.45) \times 10^{-6}$ \cr
\enddata
\vspace{-0.3in} \tablecomments{We calculate these luminosity functions 
by weighting each galaxy by the appropriate value of
$1/f(\mu_{50,r})$, corresponding to luminosity function (2) in Section
\ref{sbcorr}.}
\end{deluxetable}

\begin{deluxetable}{ccccccc}
\tablewidth{0pt}
\tablecolumns{7}
\tablecaption{\label{lf_cfact_sch} Double Schechter fits to 
luminosity functions}
\tablehead{ band & version & $M_{\ast} - 5\log_{10} h$ &
  $\phi_{\ast,1}$& $\alpha_1$ &
$\phi_{\ast,2}$& $\alpha_2$\cr
& & & $(10^{-2}$ $h^3$ Mpc$^{-3})$& & $(10^{-2}$ $h^3$ Mpc$^{-3})$ & }
\startdata
\graybox{$u$}&\graybox{corrected, $\mu_{50,r} < 24$}&\graybox{$-17.47 \pm 0.03$}&\graybox{$1.97 \pm 0.13$}&\graybox{$0.45 \pm 0.10$}&\graybox{$2.10 \pm 0.12$}&\graybox{$-1.45 \pm 0.02$}\cr
 --- & raw, $\mu_{50,r} < 24$ & $-17.59 \pm 0.04$ & $1.68 \pm 0.11$ & $0.27 \pm 0.11$ & $2.02 \pm 0.13$ & $-1.39 \pm 0.02$ \cr
\graybox{$g$}&\graybox{corrected, $\mu_{50,r} < 24$}&\graybox{$-19.22 \pm 0.05$}&\graybox{$1.29 \pm 0.08$}&\graybox{$0.14 \pm 0.13$}&\graybox{$1.18 \pm 0.08$}&\graybox{$-1.40 \pm 0.01$}\cr
 --- & raw, $\mu_{50,r} < 24$ & $-19.24 \pm 0.05$ & $1.10 \pm 0.08$ & $0.14 \pm 0.14$ & $1.27 \pm 0.09$ & $-1.34 \pm 0.02$ \cr
$r$ & ``total'' & $-20.04 \pm 0.03$ & $1.56 \pm 0.05$ & $-0.17 \pm 0.07$ & $0.62 \pm 0.04$ & $-1.52 \pm 0.01$ \cr
\graybox{---}&\graybox{corrected, $\mu_{50,r} < 24$}&\graybox{$-19.99 \pm 0.04$}&\graybox{$1.34 \pm 0.05$}&\graybox{$0.03 \pm 0.08$}&\graybox{$0.86 \pm 0.04$}&\graybox{$-1.40 \pm 0.01$}\cr
--- & raw, $\mu_{50,r} < 24$ & $-20.01 \pm 0.04$ & $1.26 \pm 0.05$ & $0.01 \pm 0.09$ & $0.86 \pm 0.05$ & $-1.34 \pm 0.01$ \cr
\graybox{$i$}&\graybox{corrected, $\mu_{50,r} < 24$}&\graybox{$-20.47 \pm 0.04$}&\graybox{$1.35 \pm 0.05$}&\graybox{$-0.18 \pm 0.08$}&\graybox{$0.59 \pm 0.04$}&\graybox{$-1.49 \pm 0.01$}\cr
 --- & raw, $\mu_{50,r} < 24$ & $-20.47 \pm 0.04$ & $1.26 \pm 0.05$ & $-0.15 \pm 0.09$ & $0.62 \pm 0.04$ & $-1.43 \pm 0.01$ \cr
\graybox{$z$}&\graybox{corrected, $\mu_{50,r} < 24$}&\graybox{$-20.68 \pm 0.04$}&\graybox{$1.50 \pm 0.06$}&\graybox{$-0.34 \pm 0.09$}&\graybox{$0.38 \pm 0.05$}&\graybox{$-1.60 \pm 0.02$}\cr
 --- & raw, $\mu_{50,r} < 24$ & $-20.67 \pm 0.04$ & $1.41 \pm 0.06$ & $-0.30 \pm 0.10$ & $0.42 \pm 0.05$ & $-1.54 \pm 0.02$ \cr
\enddata
\end{deluxetable}

\end{document}